\documentclass[aps,prd,showpacs,amsmath,amssymb]{revtex4}
\usepackage{graphicx}
\usepackage{amsfonts,bm}
\usepackage[english]{babel}

\def\be{\begin{equation}}
\def\te{\end{equation}} 
\def\ee{\end{equation}}
\def\ba{\begin{eqnarray}}
\def\bea{\begin{eqnarray}}
\def\nn{\nonumber\\}
\def\tea{\end{eqnarray}}
\def\ea{\end{eqnarray}}
\def\eea{\end{eqnarray}}

\begin{document}
\title{A hydrodynamic approach to boost invariant free streaming}

\author{E. Calzetta}
\email{calzetta@df.uba.ar}
\affiliation{Departamento de F\'isica, Facultad de Ciencias Exactas y Naturales, Universidad de Buenos Aires and IFIBA, CONICET, Cuidad Universitaria, Buenos Aires 1428, Argentina}


\begin{abstract}
We consider a family of exact boost invariant solutions of the transport equation for free streaming massless particles, where the one particle distribution function is defined in terms of a function of a single variable. The evolution of second and third moments of the one particle distribution function (the second moment being the energy momentum tensor (EMT) and the third moment the non equilibrium current (NEC)) depends only on two moments of that function. Given those two moments we show how to build a non linear hydrodynamic theory which reproduces the early time evolution of the EMT and the NEC.  The structure of these theories may give insight on nonlinear hydrodynamic phenomena on short time scales. 
\end{abstract}
\maketitle

\section{Introduction}
The goal of this paper is to build a consistent, fully nonlinear conformal hydrodynamics, capable of describing the early time evolution of a free streaming anisotropic flow while reducing to second order hydrodynamics for small deviations from ideal behavior. To avoid the proliferation of transport coefficients that plagues attempts to pursue the usual schemes, such as Chapman-Enskog or Grad expansions, to higher orders, we shall impose the strong principle that the dynamics of the theory is given by the conservation laws for the energy - momentum tensor (EMT) and a new so-called nonequilibrium current (NEC) defined through a third order  tensor. We shall impose the further restriction that both currents may be derived as gradients of a vector Massieu function. Our objective is not to find a realistic theory, but a model with the right conservation laws and a minimum of free parameters, which may be used to explore strongly nonlinear behavior in the relativistic regime.
 
The mounting evidence that relativistic heavy ion collisions \cite{BRAHMS05,PHOBOS05,STAR05,PHENIX05,Vogt07,SSS10} provide an experimental realization of relativistic real fluids \cite{Isr88} has spurred a strong interest on the characterization of those systems, and as a result we now have a fairly robust model of their dynamics in the limit of short relaxation times \cite{BJOR83,Ruus86,Risc98,CalHu08,HKB08,Roma09,HiNa12,GJS13,Huo13,Hir14,Cal13a,JeHe15}. However, this very success highlights our lack of understanding of how such a relativistic fluid comes into being in the first place \cite{Str13,Gel13,EpGe14}. To tackle that issue we need to consider phenomena on time scales which are no longer very large compared with the relaxation time, of which the non linear evolution of plasma instabilities \cite{Mrow94,Mrow07,MaMr06,SSGT06,MrTh07,MaMa07,ARS13,AKLN14,PRC13b}, shocks  \cite{OlsHis90,JoPa91,DKKM08,Bou10a,Bou10b,KKM10,KKM11,Bou14}, Kolmogorov \cite{FlWi11,Fuk13} and wave turbulence \cite{Khac08,CaRe10}  are paramount. Maybe contrary to expectations, it has been found that the language of hydrodynamics is still useful in this regime \cite{MaSt08,MaSt10}.

The assumption of free streaming, namely an infinite relaxation time, makes sense at early times. Moreover, in this limit we have simple exact solutions to the kinetic theory (reduced to just the Liouville equation for the one particle distribution function (1PDF)) which provide a test bench for hydrodynamics. A particular set of solutions which has received much attention are boost invariant, axially symmetric 1PDFs described by a function of a single variable and parametrized by a few (Milne) time dependent parameters \cite{FRS13a}. In this case it has been shown that energy momentum conservation and a second conservation law derived from the third moment of the Liouville equation determine the time evolution of those parameters, matching exactly the evolution prescribed by the full Liouville equation. This observation has led to the development of so-called anisotropic hydrodynamics \cite{MaSt10b,MaSt11,Str12,FMRS13,FRS13b,Str14,NRS14,DFRS14,Str14b}. Other exact solutions are discussed in refs. \cite{HNX14,HNX14b,NRS15,HMX15}. 

Free streaming is a worst case scenario for any relativistic hydrodynamics formalism. Collisions will tend to make the system evolve first towards a local equilibrium configuration, and then will damp out temperature gradients. Close enough to thermality, most formalisms will agree with second - order relativistic hydrodynamics; for the case of the ``divergence type'' theories we shall discuss below, we show this in Section V. The problems show up in the opposite limit, where there is no bound on how far flow can get from a thermal state. Therefore, any contending formalism must prove its worth by showing it is capable of handling free streaming.

If one considers (color) perturbations of a boost invariant background, then the usual plasma instabilities are found. The formalism of anisotropic hydrodynamics may be used to follow those instabilities and to estimate at what time the linearized theory breaks down \cite{RSA08,RS10,IRS11,ARS13b}. However, to the present author's understanding it cannot be pursued beyond that point because we do not have a full nonlinear theory whereby we could compute the back reaction of the unstable modes on the expanding background. The goal of this paper is to investigate which shape such a nonlinear hydrodynamics may have (see also \cite{{MNDLJG13}}).

To this end we shall consider theories patterned on Geroch - Lindblom ``divergence type'' hydrodynamics (DTT)\cite{GeLi90,GeLi91,Cal98,CaTh01,PRC09,PRC10c,PRC10b}. The equations of motion are given by the conservation laws for the EMT $T^{\mu\nu}$ and the NEC $A^{\mu\nu\rho}$ \cite{Lax73}. We shall consider a conformal colorless fluid,  so there will be no conservation law for charge or particle number. The degrees of freedom of the theory are a inverse temperature four vector $\beta^{\mu}$ and a second order nonequilibrium tensor $\zeta^{\mu\nu}$. A most important assumption is that there is a Massieu function current $\Phi^{\mu}$ such that  $T^{\mu\nu}$ and  $A^{\mu\nu\rho}$ may be obtained as derivatives of  $\Phi^{\mu}$ with respect to $\beta^{\mu}$ and $\zeta^{\mu\nu}$. This assumption is key to give the approach predictive power. The hydrodynamic formalism is linked to kinetic theory by relating  $T^{\mu\nu}$ and  $A^{\mu\nu\rho}$ to the second and third moments of the 1PDF \cite{NaRe95,DKR10,DMNR12,PRC10a,PRC13a}.

We shall match those theories against a family of exact boost invariant solutions of the Liouville equations. For these solutions,  $T^{\mu\nu}$ and  $A^{\mu\nu\rho}$ are determined by just two moments of the 1PDF. Given those two parameters, we show how  a fully nonlinear hydrodynamic theory may be build that reproduces the known early time evolution of  $T^{\mu\nu}$ and  $A^{\mu\nu\rho}$. The resulting theories are probably too complex to be relevant in practice. However, we also show a simple theory which captures the early time behavior, and may  be useful to investigate nonlinear effects in this regime.

The rest of the paper is organized as follows. In next section we present the boost invariant, axisymmetric free streaming solutions and compute the EMT and the NEC. Then in Section III we derive the most general hydrodynamic theory where these currents may be derived as derivatives of a single vector potential. Finally in Section IV we match hydrodynamics to the exact solution, and in Section V we analyze the inclusion of viscosity and the behavior of the theory for small deviations from ideal behavior, whereby it reduces to a version of second order hydrodynamics \cite{DNMR12,FJMRS15}.

We conclude with some brief remarks. Some technical details concerning the derivation of eq. (\ref{gen1}) below are given in the Appendix.

\section{Boost Invariant Free Streaming}
We consider a situation where the particles live in $3+1$ dimensions but the 1PDF is independent of the ``transverse'' coordinates $x$ and $y$. For simplicity we shall assume axial symmetry in the transverse plane. We will use both cartesian coordinates with interval

\be 
ds^ 2=-dt^ 2+dz^2+dx^ 2+dy^2
\te 
and Milne coordinates $t=\tau \cosh\chi$, $z=\tau \sinh\chi$, whereby

\be 
ds^ 2=-d\tau^ 2+\tau^2d\chi^2+dx^ 2+dy^2
\te 
In the absence of collisions the transport equation reduces to Liouville's 

\be 
\left\lbrace p^0\frac{\partial}{\partial t}+p^z\frac{\partial}{\partial z}\right\rbrace f_B=0 
\te 
We consider a massless theory, so 

\be 
p^0=\sqrt{p^{z2}+p_{\perp}^2}
\te 
The most general solution of the transport equation is 

\be 
f_B=f_0\left[ p^z t-p^0z,\mathbf{p_{\perp}}\right] 
\te
where $f_0$ may be any function. Introducing the rapidity $Y$ through $p^0=p_{\perp}\cosh Y$, $p^z=p_{\perp}\sinh Y$ we see that 

\be 
f_B=f_0\left[ \tau p_{\perp}\sinh\left( Y-\chi\right) ,\mathbf{p_{\perp}}\right] 
\te 
On the other hand, the relationships 

\bea 
p^0&=&\cosh\chi\; p^{\tau}+\tau\sinh\chi\; p^{\chi}\nn
p^z&=&\sinh\chi\; p^{\tau}+\tau\cosh\chi\; p^{\chi}
\tea 
may be inverted to yield 

\bea 
p^{\tau}&=&p_{\perp}\cosh\left( Y-\chi\right)\nn 
p^{\chi}&=&\frac{p_{\perp}}{\tau}\sinh\left( Y-\chi\right)
\tea 
Therefore we see that the general solution to the Liouville equation is 

\be 
f_B=f_0\left[ \tau^2 p^{\chi} ,\mathbf{p_{\perp}}\right]
\label{generic}
\te 
and is boost invariant if we adopt $\mathbf{p_{\perp}}$ and $p^{\chi}$ as independent variables. Incidentally, observe that a thermal 1PDF is not a solution of the Liouville equation. 

\subsection{Moments of the free streaming 1PDF}
We consider a solution of the type eq. (\ref{generic}), we only assume $f_0$ is even in $p^{\chi}$ and axially symmetric in the transverse plane. The covariant moments are

\be
A_n^{\mu_1,...\mu_n}=\tau\int\frac{dp^{\chi}d^2\mathbf{p_{\perp}}}{\left(2\pi\right)^3p^0}p^{\mu_1}....p^{\mu_n}f_0
\te
They are totally symmetric, traceless on any two indexes, and obey conservation laws

\be
\nabla_{\mu_n}A_n^{\mu_1,...\mu_n}=I_{n-1}^{\mu_1,...\mu_{n-1}}
\label{conslaw}
\te
where

\be
I_n^{\mu_1,...\mu_n}=\frac{\tau}{\tau_R}\int\frac{dp^{\chi}d^2\mathbf{p_{\perp}}}{\left(2\pi\right)^3p^0}p^{\mu_1}....p^{\mu_n}I_{col}\left[f_0\right]
\te
We are interested in $A_2^{\mu\nu}=T^{\mu\nu}$, $A_3^{\mu\nu\rho}=A^{\mu\nu\rho}$ and their sources $I_1^{\mu}=0$ and $I_2^{\mu\nu}=I^{\mu\nu}$. For simplicity we shall call the former the energy-momentum tensor (EMT) and the latter the non equilibrium current (NEC). For the boost invariant symmetric solution the only nonzero components of the EMT are $T^{\chi}_{\chi}=T_{\chi}$, $T^{ab}=\delta^{ab}T_T$ and $T^{00}=T_{\chi}+2T_T$. The only nontrivial Christoffel symbols are $\Gamma^{\tau}_{\chi\chi}=\tau$ and $\Gamma^{\chi}_{\tau\chi}={\tau}^{-1}$. The conservation law implies

\be
\frac 1{2\tau}\frac d{d\tau}\left(\tau^2 T^{00}\right)= T_T
\label{tcons}
\te
The nontrivial third moments are $A^{0\chi\chi}$, $A^{0ab}=\delta^{ab}A_T$, $A^{000}=\tau^2A^{0\chi\chi}+2A_T$ and their permutations. We now have

\bea
A_T&=&\frac{\tau}2\int\frac{dp^{\chi}d^2\mathbf{p_{\perp}}}{\left(2\pi\right)^3}p_{\perp}^2f_0\left[ \tau^2 p^{\chi} ,\mathbf{p_{\perp}}\right]\nn
&=&\frac1{2\tau}\int\frac{dPd^2\mathbf{p_{\perp}}}{\left(2\pi\right)^3}p_{\perp}^2f_0\left[ P,\mathbf{p_{\perp}}\right]=\frac1{\tau}A_{T0}
\tea

\bea
A^{0\chi\chi}&=&\tau\int\frac{dp^{\chi}d^2\mathbf{p_{\perp}}}{\left(2\pi\right)^3}p^{\chi 2}f_0\left[ \tau^2 p^{\chi} ,\mathbf{p_{\perp}}\right]\nn
&=&\frac1{\tau^5}\int\frac{dPd^2\mathbf{p_{\perp}}}{\left(2\pi\right)^3}P^2f_0\left[ P,\mathbf{p_{\perp}}\right]=\frac1{\tau^5}A^{0\chi\chi}_0
\label{asymp}
\tea
These results can also be derived from the conservation law

\be
A^{\mu\nu\rho}_{\;\;\;;\mu}=\frac1{\tau}\frac d{d\tau}\left(\tau A^{0\nu\rho}\right)+\Gamma^{\nu}_{\mu\lambda}A^{\mu\lambda\rho}+\Gamma^{\rho}_{\mu\lambda}A^{\mu\nu\lambda}=0
\label{cons}
\te
which follows from $I_2^{\mu\nu}=0$ in eq. (\ref{conslaw}) for free streaming. $A_T\approx\tau^{-1}$ is evident. Setting $\nu=\rho=1$ we get

\be
\frac1{\tau}\frac d{d\tau}\left(\tau A^{0\chi\chi}\right)+\frac 4{\tau}A^{0\chi\chi}=0
\te
As expected. Finally, $\nu=\rho=0$ leads to

\be
\frac1{\tau}\frac d{d\tau}\left(\tau A^{000}\right)+2{\tau}A^{0\chi\chi}=0
\te 

\subsection{A restricted solution}

The single restriction of solving the Liouville equation still leaves a very large scope for the 1PDF. To be able to make progress, we shall restrict our discussion to a particular set of 1PDFs, which are obtained from an isotropic PDF by an anisotropic rescaling. This family of solutions was introduced by Romatschke and Strickland \cite{RomStr03}, and has played a large role in the anisotropic hydrodynamics literature \cite{MaSt10b,MaSt11,Str12,FMRS13,FRS13b,Str14,NRS14,DFRS14,Str14b}.

\begin{center}
\begin{figure}[htb]
\scalebox{0.47}{\includegraphics{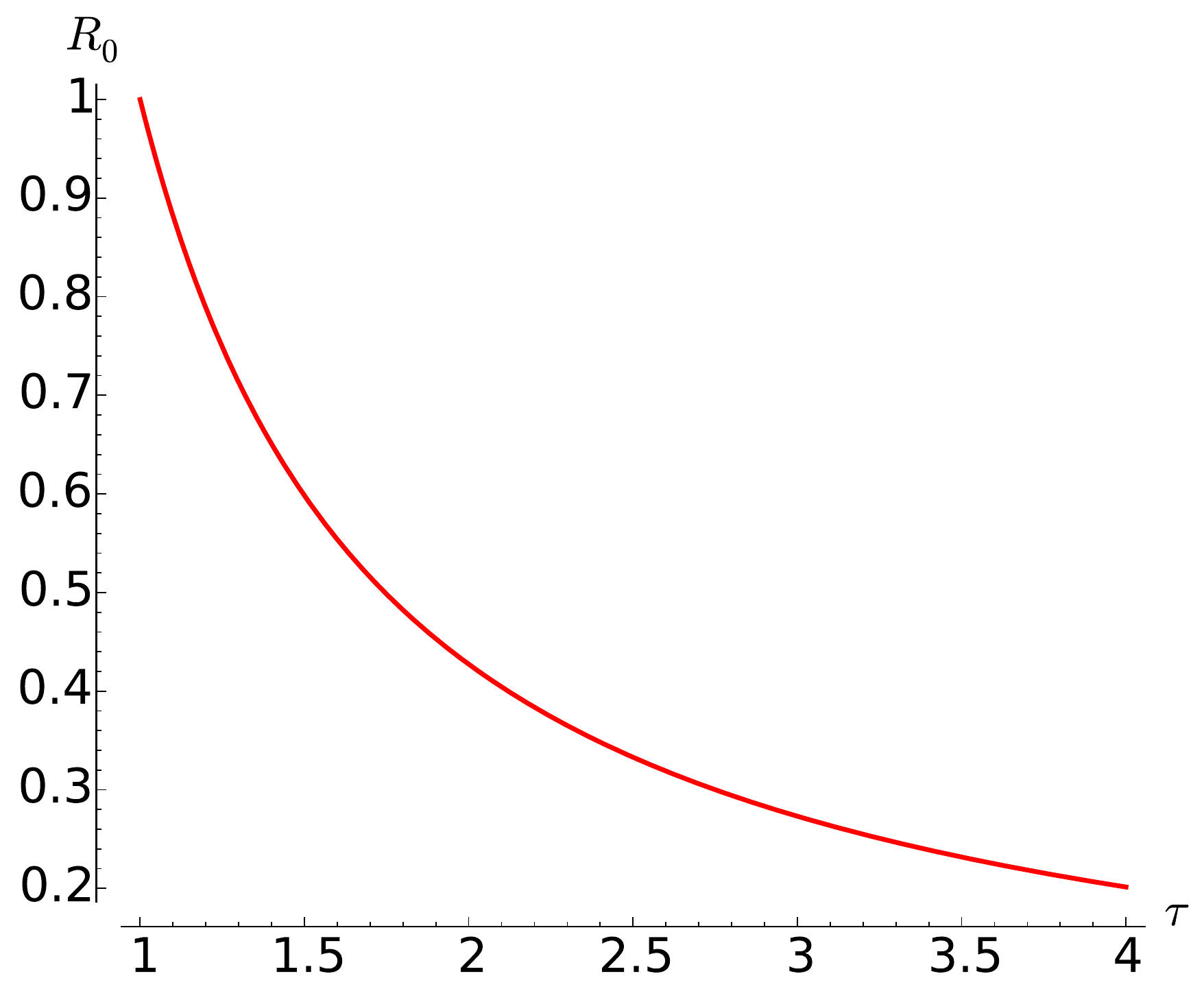}}
\caption{(Color online) $R_0$ as a function of $\tau$. }
\label{f0plot}
\end{figure}
\end{center}

Concretely we assume the 1PDF is a function of the single variable

\be
\Xi^2=C_0p_{\perp}^2+C_1\tau^4p^{\chi 2}
\te
where the $C_i$ are constants ($C_1$ is dimensionless, while $C_0$ has units of inverse  temperature squared). We wish to compute momenta of the distribution function, which are of the form

\be
\left\langle T_n\right\rangle={2\tau}\int_0^{\infty}dp^{\chi}\int\frac{d^2\mathbf{p_{\perp}}}{\left(2\pi\right)^3\sqrt{\tau^2p^{\chi 2}+p_{\perp}^2}}T_n\;f_0\left[ C_1\tau^4 p^{\chi 2} +C_0{p_{\perp}}^2\right]
\te
It is therefore convenient to introduce new variables

\bea
\frac{\phi}{C_0}&=& c\tau^4 p^{\chi 2} +{p_{\perp}}^2\nn
\psi &=&\tau^2p^{\chi 2}+p_{\perp}^2
\tea
$c=C_1/C_0$. Then

\bea
p^{\chi 2}&=&\frac{\frac{\phi}{C_0}-\psi}{\tau^2\left(c\tau^2-1\right)}\nn
p_{\perp}^2 &=&\frac{c\tau^2\psi-\frac{\phi}{C_0}}{\left(c\tau^2-1\right)}
\tea
so

\be
dp^{\chi}d^2\mathbf{p_{\perp}}=\frac{\pi}{2p^{\chi}}dp^{\chi 2}d^2\mathbf{p_{\perp}}=\frac{\pi}{2C_0\tau\sqrt{c\tau^2-1}}\frac{d\phi d\psi}{\sqrt{\frac{\phi}{C_0}-\psi}}
\te
Positivity of $p^{\chi 2}$ and $p_{\perp}^2$ implies 

\be
\frac{\phi}{C_0}\ge\psi\ge\frac{\phi}{C_1\tau^2}
\te
To compute $T^{00}$ we choose $T_n=\psi$. We get

\be
T^ {00}=\frac{1}{8\pi^2C_0\sqrt{c\tau^2-1}}\int \;d\phi\;f_0\left[ \phi\right] \int_{{\phi}/{C_1\tau^2}}^{{\phi}/{C_0}}d\psi\frac{\sqrt{\psi}}{\sqrt{\frac{\phi}{C_0}-\psi}}
\te
Leading to 

\be
T^ {00}=\sigma T_0^4R_0\left( \sqrt{c}\tau\right) 
\label{kint0}
\te
where $\sigma$ is the Stefan - Boltzmann constant and

\bea 
J_{a}&=&\int \;d\phi\;\phi^a f_0\left[ \phi\right] \nn
\sigma T_0^4&=&\frac{J_1}{4\pi^2C_0^2}\nn
R_0\left( x\right) &=&\frac12\left[\frac 1{x^2}+\frac{\tan^ {-1}\left[\sqrt{x^2-1} \right] }{\sqrt{x^2-1}} \right] 
\tea
We show a plot of $R_0\left( x\right) $ in fig. (\ref{f0plot})

\begin{center}
\begin{figure}[htb]
\scalebox{0.47}{\includegraphics{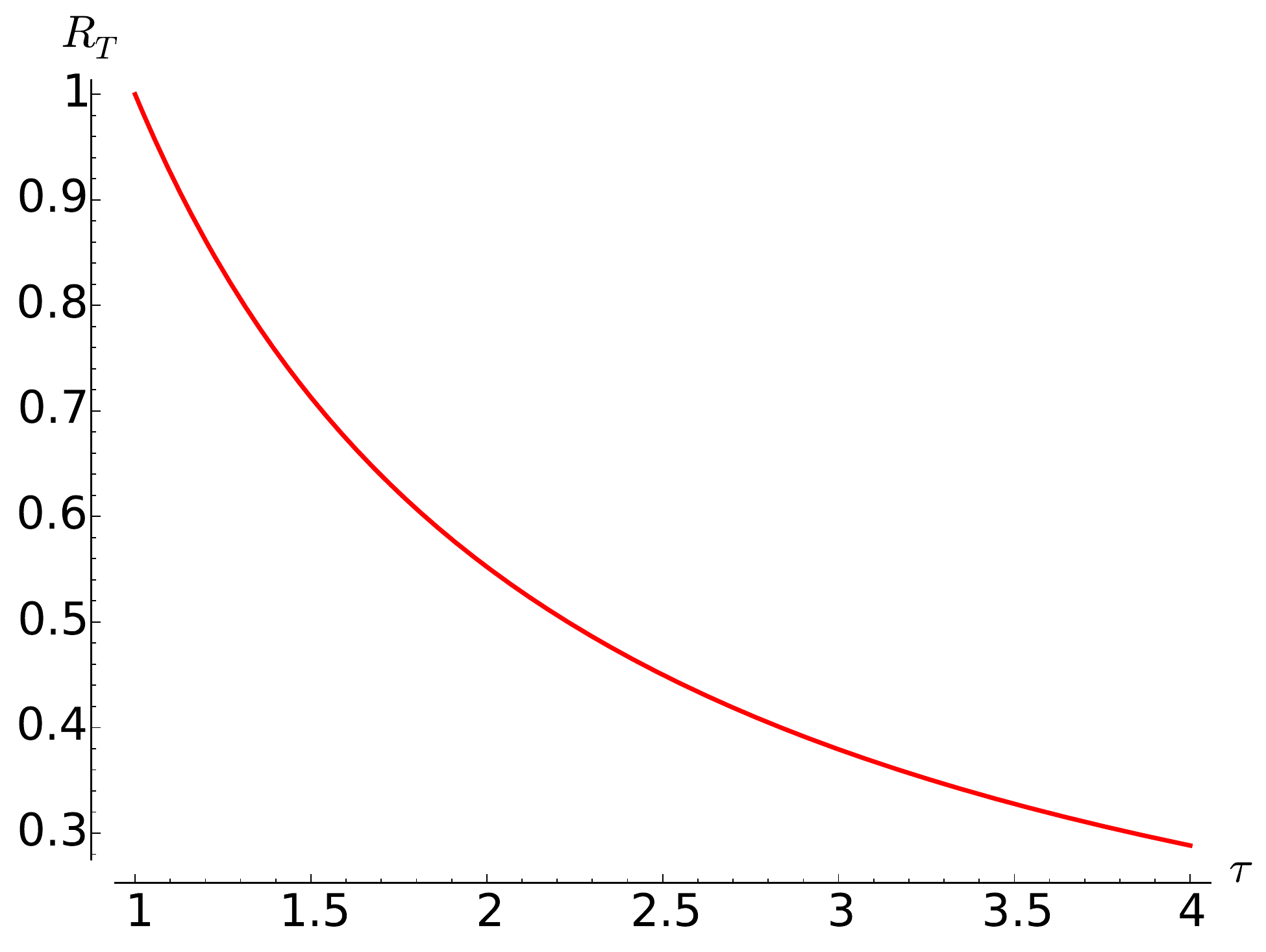}}
\caption{(Color online) $R_T$ as a function of $\tau$. }
\label{fTplot}
\end{figure}
\end{center}

To compute $T_T$ we set $T_n=p_{\perp}^2/2$, or else we may get $T_T$ from the conservation law (we write $s=c\tau^2-1$)

\be
T_T=\frac d{ds}\left(\left(s+1\right) T^{00}\right)= \frac13\sigma T_0^4R_T\left(\sqrt{c}\tau\right)  
\label{kintt}
\te

\be 
R_T\left( x\right) =\frac 3{4\left( x^2-1\right) } \left[1+\frac{\left( x^2-2\right)}{\sqrt{x^2-1}}\tan^ {-1}\left[\sqrt{x^2-1} \right]\right] 
\te
We show a plot of $R_T\left( x\right) $ in fig. (\ref{fTplot})

Once $R_0$ and $R_T$ are known, we get the longitudinal component as 

\be 
T^{\chi}_{\chi}=\frac13\sigma T_0^4R_{\chi}\left(\sqrt{c}\tau\right)  
\label{kinchi}
\te
where 

\be 
R_{\chi}=3R_0-2R_T
\te 

\begin{center}
\begin{figure}[htb]
\scalebox{0.47}{\includegraphics{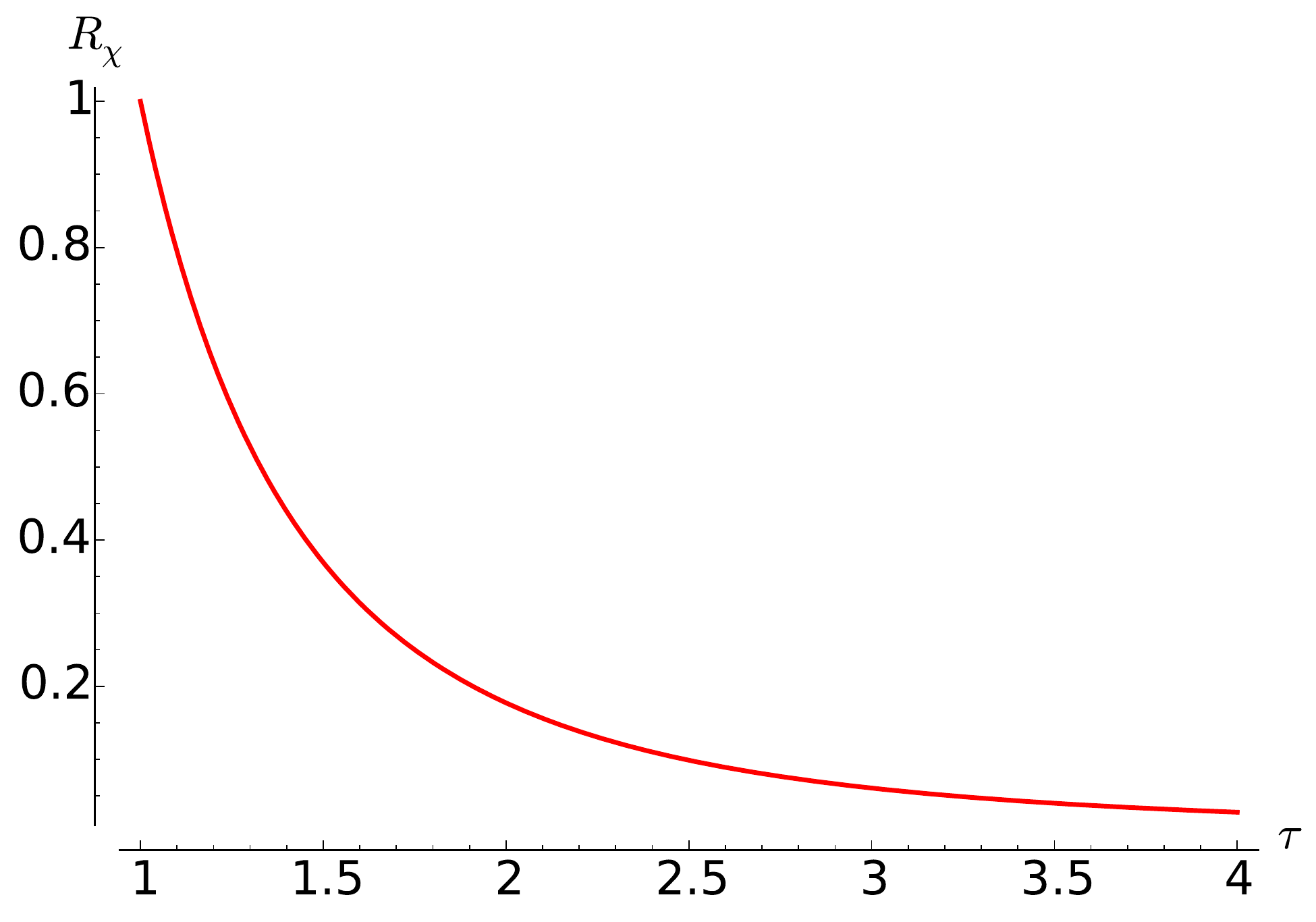}}
\caption{(Color online) $R_{\chi}$ as a function of $\tau$. }
\label{fcplot}
\end{figure}
\end{center}

We show a plot of $R_{\chi}\left( x\right) $ in fig. (\ref{fcplot})

\begin{center}
\begin{figure}[htb]
\scalebox{0.47}{\includegraphics{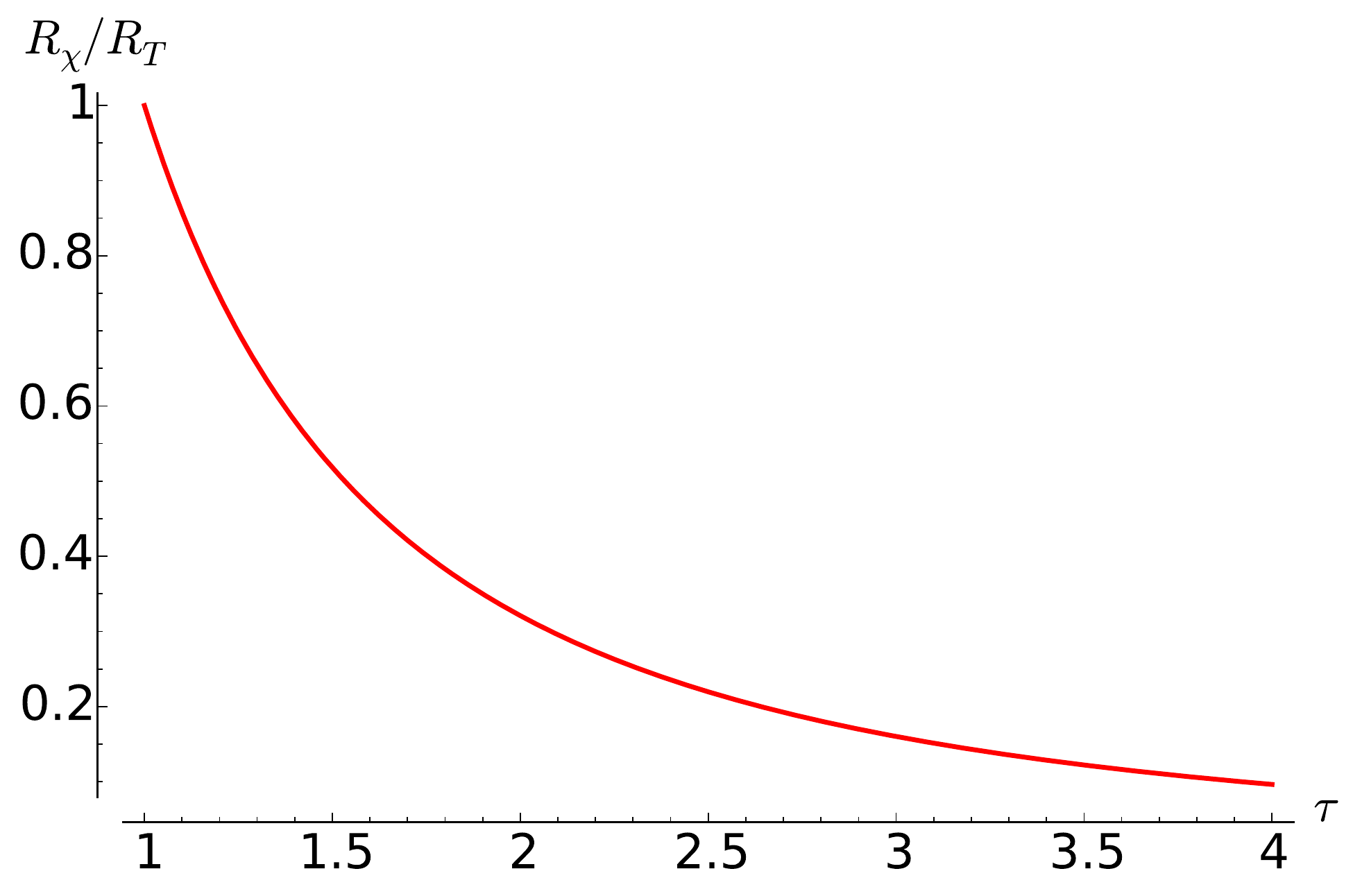}}
\caption{(Color online) The anisotropy parameter $R_{\chi}/R_T$ as a function of $\tau$. }
\label{faplot}
\end{figure}
\end{center}

The anisotropy parameter is defined as $R_{\chi}/R_T$, and decreases monotonically, as shown in fig. (\ref{faplot})

To compute $A^{0\chi\chi}=A_{\chi}/\tau^ 2$ we set $T_n=\psi^{1/2} p^{\chi 2}$

\be
A_{\chi}=\frac{1}{8\pi^2C_0\left( c\tau^2-1\right) ^{3/2}}\int \;d\phi\;f_0\left[ \phi\right] \int_{{\phi}/{C_1\tau^2}}^{{\phi}/{C_0}}d\psi\;\sqrt{\frac{\phi}{C_0}-\psi}=
\sigma T_0^5K_A\frac1{c^{3/2}\tau^3}
 \label{kinachi}
\te

\be 
\sigma T_0^5K_A=\frac{J_{3/2}}{12\pi^2C_0^{5/2}}
\te 
To compute $A_T$ we set $T_n=\psi^{1/2} p_{\perp}^2/2$

\be
A_T=\frac{1}{16\pi^2C_0\left( c\tau^2-1\right) ^{3/2}}\int \;d\phi\;f_0\left[ \phi\right] \int_{{\phi}/{C_1\tau^2}}^{{\phi}/{C_0}}d\psi\;\frac{c\tau^2\psi-\frac{\phi}{C_0}}{\sqrt{\frac{\phi}{C_0}-\psi}}=
\sigma T_0^5K_A\frac1{c^{1/2}\tau}
\label{kinat}
\te
From this point on we set $c=1$, without loss of generality.

\section{The most general (DTT-inspired) non linear conformal hydro}

We wish to see if it is possible to cast hydrodynamics into something resembling a DTT framework, with the goal to provide a fully nonlinear model. For simplicity we consider a conformal, neutral fluid.

The degrees of freedom are a four-vector $\beta_{\mu}$ and a traceless, symmetric tensor $\zeta_{\mu\nu}$. The equations of motion are of divergence type

\bea 
T^{\mu\nu}_{;\mu}&=&0\nn
A^{\mu\nu\rho}_{;\mu}&=&I^{\nu\rho}
\label{dtteq}
\tea 
$T^{\mu\nu}$ is traceless, $A^{\mu\nu\rho}$ is totally symmetric and traceless on any two indexes. We expect they are derivable from a potential

\bea 
T^{\mu\nu}&=&\frac{\partial\Phi^{\mu}}{\partial\beta_{\nu}}\nn
A^{\mu\nu\rho}&=&\frac{\partial\Phi^{\mu}}{\partial\zeta_{\nu\rho}}
\label{potential}
\tea 
The symmetry of the EMT implies that $\Phi^{\mu}$ itself is a gradient

\be 
\Phi^{\nu}=\frac{\partial\Phi}{\partial\beta_{\nu}}
\te 
Since $\Phi$ is a scalar, it can only depend on other scalars. 

The whole point of the DTT framework is to write down a consistent (in particular causal) relativistic hydrodynamics with a minimum of hypothesis (ideally, with no assumptions at all) regarding the underlying description, e. g. kinetic theory or quantum field theory. 
The reduction of the theory to the specification of a single scalar function is what gives the theory its predictive power, and therefore it is key to the appeal of this approach

Let us observe that if we wish to identify $T^{\mu\nu}$ and $A^{\mu\nu\rho}$ with the moments of the 1PDF as above, then the dimensions of $\left[ T^{\mu\nu}\right] =E^4$, where $E$ stands for energy (we use natural units with $\hbar=c=k_B=1$) and $\left[A^{\mu\nu\rho}\right]=E^5$ . With the usual identification $\beta_{\mu}=u^{\mu}/T$ with $u^{\mu}$ dimensionless (and $u^2=u^{\mu}u_{\mu}=-1$) and $T$ the temperature with $\left[ T\right] =E$, we find $\left[ \Phi^{\nu}\right] =E^3$, $\left[ \Phi\right] =E^2$ and $\left[\zeta_{\mu\nu}\right] =E^{-2}$ . This suggests introducing a dimensionless tensor variable $v_{\mu\nu}= T^2\zeta_{\mu\nu}$ and a dimensionless potential $\Phi=T^2\varphi$. 

If there are no dimensionful constants, then $\varphi$ cannot depend on $T$, but only on the dimensionless invariants $X^a=\left( v^a\right) ^{\mu}_{\mu}=T^{2a}\left( \zeta^a\right) ^{\mu}_{\mu}$ and $Y^a=u_{\mu}\left( v^a\right) ^{\mu}_{\nu}u^{\nu}=T^{2a }u_{\mu}\left( \zeta^a\right) ^{\mu}_{\nu}u^{\nu}$, where $a=1$ to $4$ (no other linearly independent invariants exist in $4$ dimensions). For definiteness, we shall require $u^{\mu}$ to be the Landau - Lifshitz hydrodynamic velocity, namely, that $T^ {\mu\nu}u_{\nu}=-\rho u^ {\mu}$. This implies  $v^ {\mu\nu}u_{\nu}=0$. Therefore that $X^1$ and the $Y^a$ vanish on shell but they may contribute to the potential derivatives. We shall not assume further that $T$ is the Landau-Lifshitz temperature, that is, we do not require $\rho=\sigma T^ 4$; $T$ is rather the temperature of a fiducial ideal fluid.

From the usual formulae

\bea
\frac{\partial T}{\partial\beta_{\mu}}&=&T^2u^{\mu}\nn
\frac{\partial u^{\nu}}{\partial\beta_{\mu}}&=&T\Delta^{\mu\nu}
\tea
$\Delta^{\mu\nu}=g^{\mu\nu}+u^{\mu}u^{\nu}$. We get $\Phi^{\mu}=T^3\varphi^{\mu}$, where

\bea 
\varphi^{\mu}&=& u^{\mu}\phi +2\sum_{a=1}^4\frac{\partial\varphi}{\partial Y^a}\left( v^a\right) ^{\mu}_{\nu}u^{\nu}\nn
\phi&=&2\left[ \varphi +\sum_{a=1}^4\left(a X^a\frac{\partial\varphi}{\partial X^a}+\left( a+1\right) Y^a\frac{\partial\varphi}{\partial Y^a}\right) \right]
\tea
Therefore, on shell

\bea
\Phi^{\mu}&=&T^3\phi u^{\mu}\nn
T^{\mu\nu}&=&T^4t^{\mu\nu}\nn
A^{\mu\nu\rho}&=&T^5a^{\mu\nu\rho}
\tea
where

\bea
\phi &=&2\left[ \varphi +\sum_{a=2}^4 aX^a\frac{\partial\varphi}{\partial X^a}\right]\nn
t^{\mu\nu}&=&\left[ u^{\mu}u^{\nu}+\frac13\Delta^{\mu\nu}\right] \rho +2{v}^{\mu\nu}\left[ \frac{\partial\varphi}{\partial Y^1}+\frac12X^2\frac{\partial\varphi}{\partial Y^3}+\frac13X^3\frac{\partial\varphi}{\partial Y^4}\right] \nn
&+&2\left(\tilde{v}^2\right)^{\mu\nu}\left[ \frac{\partial\varphi}{\partial Y^2}+\frac12X^2\frac{\partial\varphi}{\partial Y^4}\right]\nn
a^{\mu\nu\rho}&=&\left[ 3u^{\mu}u^{\nu}u^{\rho}+u^{\mu}\Delta^{\nu\rho}+\Delta^{\mu\nu}u^{\rho}+\Delta^{\mu\rho}u^{\nu}\right] \left[\frac{\partial\varphi}{\partial Y^1}+ \sum_{a=3}^4\frac{\partial\varphi}{\partial Y^a}\frac{X^{a-1}}{3}\right]\nn
&+&\left[ \frac{\partial\varphi}{\partial Y^2}+\frac12X^2\frac{\partial\varphi}{\partial Y^4}\right] \left(v^{\mu\nu}u^{\rho}+v^{\mu\rho}u^{\nu}+v^{\nu\rho}u^{\mu}\right)\nn
&+&\frac{\partial\varphi}{\partial Y^3}\left(\left( \tilde{v}^{2}\right) ^{\mu\nu}u^{\rho}+\left( \tilde{v}^{2}\right) ^{\mu\rho}u^{\nu}+\left( \tilde{v}^{2}\right) ^{\nu\rho}u^{\mu}\right)\nn
\label{gen1}
\tea
here

\bea
\left(\tilde{v}^a\right)^{\mu\nu}&=&\left( v^a\right)^{\mu\nu}-\frac{X^a}3\Delta^{\mu\nu}\nn
\rho&=&6\varphi+6\sum_{a=2}^4 aX^a\frac{\partial\varphi}{\partial X^a} +2\sum_{a=2}^4\frac{\partial\varphi}{\partial Y^a}X^{a}\nn
&=&3\phi +2\sum_{a=2}^4\frac{\partial\varphi}{\partial Y^a}X^{a}
\label{gen2}
\tea
We give a detailed derivation of these equations in the Appendix.
Equations (\ref{gen1}) and (\ref{gen2}) are the most general expressions for the EMT and NEC for conformal theories derived from a potential. They are the most important result of this paper, because they display in full the inner relations between the transport functions appearing in one and the other. Of course, for boost invariant, axisymmetric flows, these equations take a much simpler form, which we derive presently.

In general $v^{\mu}_{\nu}=\mathrm{diag}\left(0,-2v_+,v_++v_-,v_+-v_-\right)$. If moreover the solution is axially symmetric, then $v_-=0$. Writing $v_+=v$ we have $X^2=6v^2$, $X^3=-6v^3$, $X^4=18v^4$ and $\left( \tilde{v}^2\right)^{\mu\nu}=-vv^{\mu\nu}$. We then get, calling

\bea
F&=&\frac{\partial\varphi}{\partial Y^1}+2v^2\frac{\partial\varphi}{\partial Y^3}-2v^3\frac{\partial\varphi}{\partial Y^4}\nn
G&=&\frac{\partial\varphi}{\partial Y^2}-v\frac{\partial\varphi}{\partial Y^3}+3v^2\frac{\partial\varphi}{\partial Y^4}
\tea

\bea
a^{\mu\nu\rho}&=&F\left( 3u^{\mu}u^{\nu}u^{\rho}+u^{\mu}\Delta^{\nu\rho}+\Delta^{\mu\nu}u^{\rho}+\Delta^{\mu\rho}u^{\nu}\right) \nn
&+&G \left(v^{\mu\nu}u^{\rho}+v^{\mu\rho}u^{\nu}+v^{\nu\rho}u^{\mu}\right)\nn
t^{\mu\nu}&=&\left[ u^{\mu}u^{\nu}+\frac13\Delta^{\mu\nu}\right] \rho +2{v}^{\mu\nu}\left[ F-vG\right]\nn
\rho &=&3\phi +12v^2G
\tea
The functions $\rho$, $F$ and $G$ are not independent but must satisfy the constraints given in the Appendix, eqs. (\ref{const1}, \ref{const2}, \ref{const3}, \ref{const4}) and (\ref{const5}). 

\section{Matching to the free streaming solution}

We now will seek one theory within the family identified above matching the hydrodynamic currents of the free streaming solution. There are four currents to match, namely $A_T$ (eq. (\ref{kinat})), $A_{\chi}$ (eq. (\ref{kinachi})), $T_T$ (eq. (\ref{kintt})) and $T^{00}$ (eq. (\ref{kint0})) or equivalently $T_{\chi}=T^{00}-2T_T$.
At $\tau=1$ the anisotropy parameter $\alpha =T_{\chi}/T_T=1$ and also $A_{\chi}=A_T$. It is natural to identify this point as $v=0$, $T=T_0$. We also write  $F=\sigma K_Af\left(v\right)$, $G=\sigma K_Ag\left(v\right)$ and $T=tT_0$. The matching conditions are

\bea
t^5\left( f+ vg\right) &=&\frac 1{\tau}\nn
t^5\left( f-2 vg\right) &=&\frac 1{\tau^3}\nn
 \rho t^4&=&\sigma R_0\left( \tau\right) \nn
 2K_At^4v\left( f-vg\right) &=&\frac{-1}{3}\left[ R_0\left( \tau\right)-R_T\left( \tau\right)\right] 
\tea

\begin{center}
\begin{figure}[htb]
\scalebox{0.47}{\includegraphics{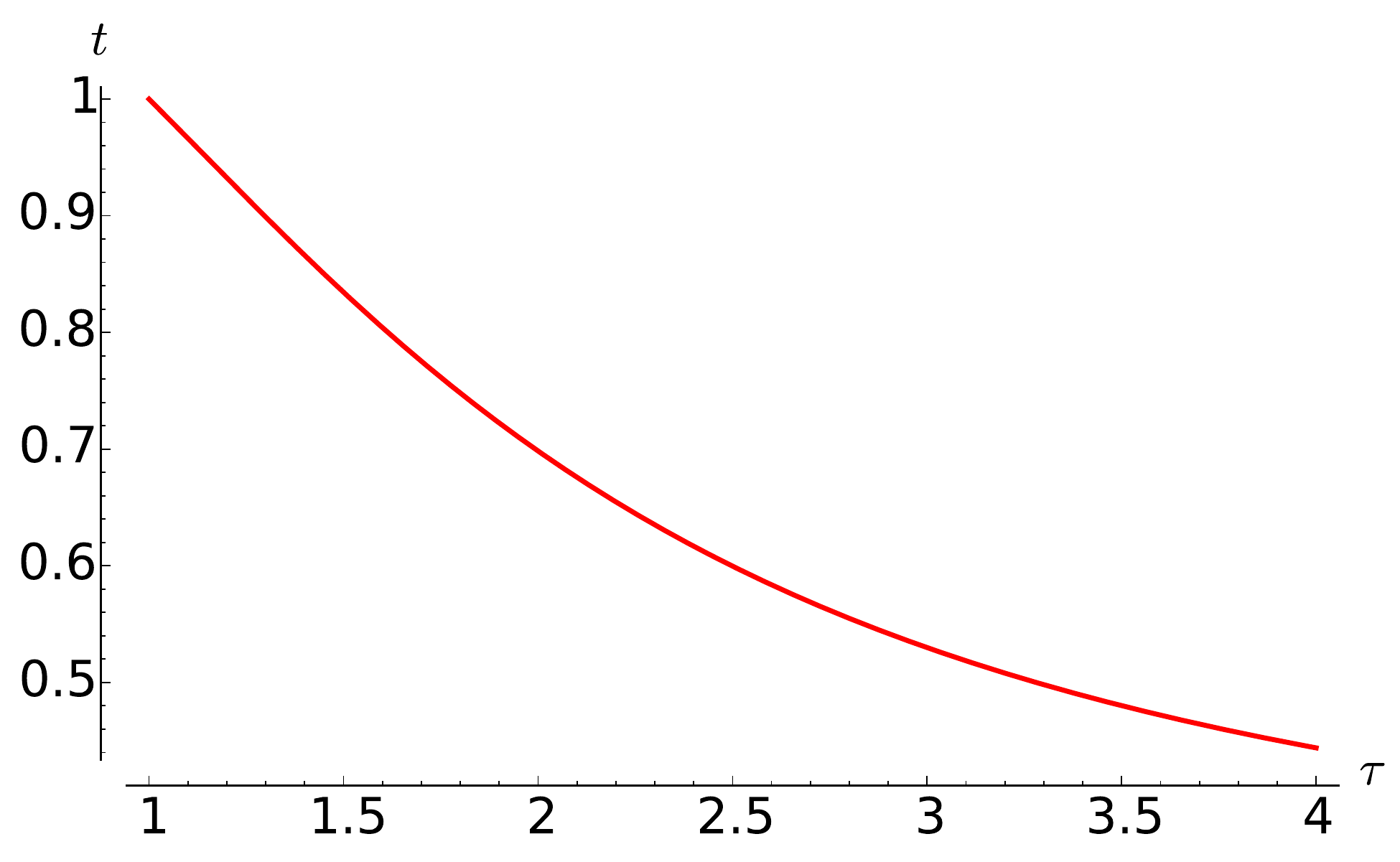}}
\caption{(Color online) The dimensionless temperature parameter $t$  as a function of time  }
\label{plott}
\end{figure}
\end{center}

\begin{center}
\begin{figure}[htb]
\scalebox{0.47}{\includegraphics{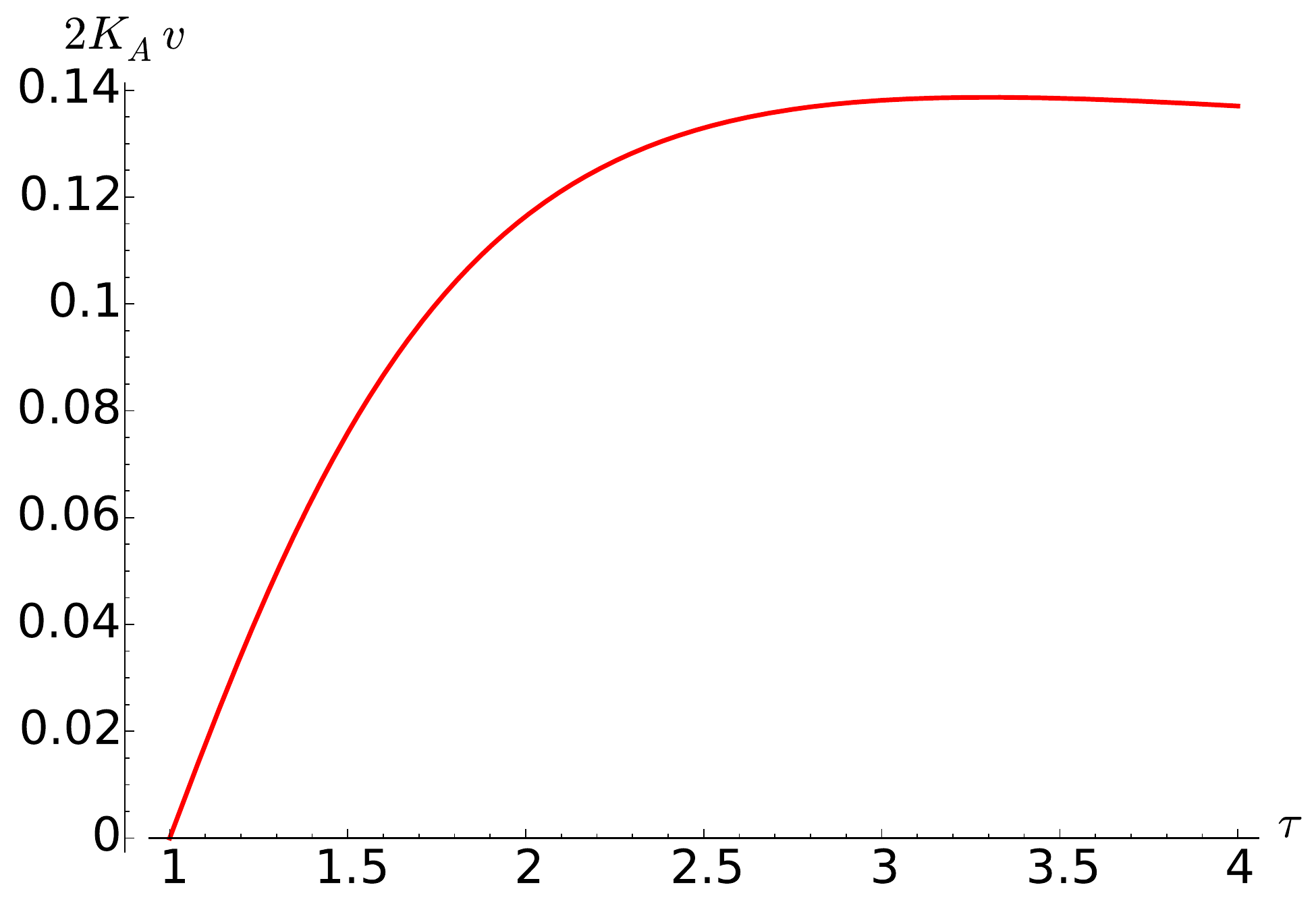}}
\caption{(Color online) The dimensionless nonideal parameter $2K_Av$ as a function of time  }
\label{plotv}
\end{figure}
\end{center}

Besides $T^{\mu\nu}$ and $A^{\mu\nu\rho}$ we can build the conserved vector

\be 
S^{\mu}=\Phi^{\mu}-\beta_{\nu}T^{\mu\nu}-\zeta_{\rho\sigma}A^{\mu\rho\sigma}
\label{entropy}
\te 
$S^ {\mu}$ represents the entropy density. On shell $S^ {\mu}=\left( 4/3\right) \sigma T_0^3s u^{\mu}$, where (see eqs. (\ref{gen1}) and (\ref{gen2})) 

\be 
s=\frac{3}{4}t^3\left[\sigma^{-1} \left( \phi +\rho\right)  -6K_Agv^2\right] =t^3\left[ \frac{\rho}{\sigma}-\frac{15}{2}K_Av^2g\right] 
\te
Conservation implies $s=1/\tau$, and so we get a fifth matching condition

\be 
\frac{R_0\left( \tau\right) }{t^4}=\frac 1{t^ 3\tau}+\frac{15}{2}K_Av^2g
\te
These equations may be solved as follows. Solving the first two we find

\bea 
g&=&\frac{\tau^2-1}{3vt^5\tau^3}\nn
f&=&\frac{2\tau^2+1}{3t^5\tau^3}
\tea 
so $f-vg=\left(\tau^2+2\right)/3\tau^3 t^ 5$ and the fourth condition gives

\be 
\frac{2K_Av}{t}=\frac{\tau^3}{\left(\tau^2+2\right)}\left[ R_T\left( \tau\right)-R_0\left( \tau\right)\right] 
\label{vtt}
\te 
whereby the fifth condition yields

\be 
t=\tau\left\lbrace R_0\left( \tau\right)-\frac 54\left( \frac {\tau^2-1}{\tau^2+2}\right) \left[ R_T\left( \tau\right)-R_0\left( \tau\right)\right] \right\rbrace
\label{temp}
\te
We plot these solutions in figs. (\ref{plott}) and (\ref{plotv}).

\begin{center}
\begin{figure}[htb]
\scalebox{0.47}{\includegraphics{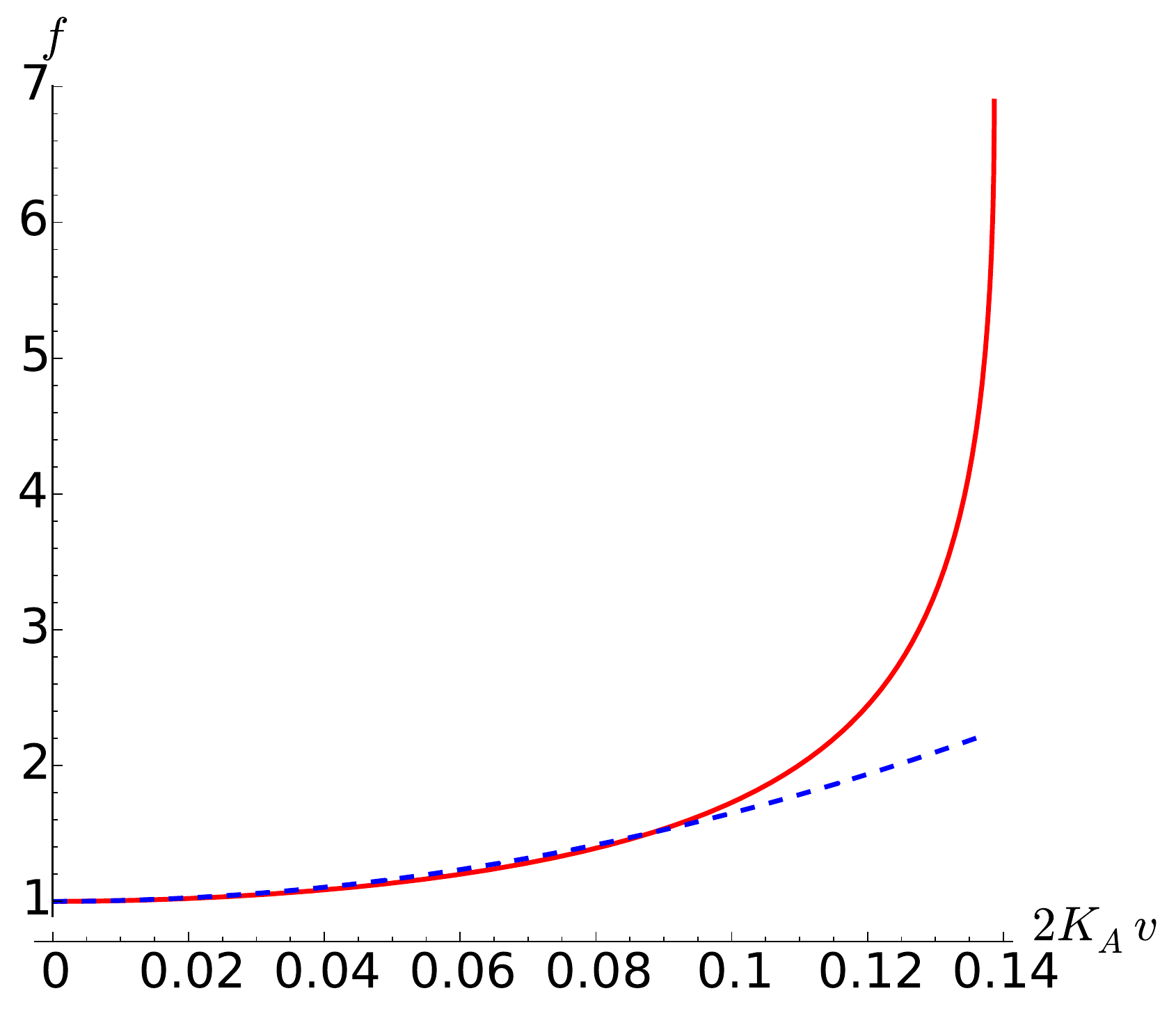}}
\caption{(Color online) Full line: the exact function $f=f\left( v\right) $; dashes: the approximation to $f$ from eq. (\ref{aproxfg}) both as a function of the invariant $2K_Av$  }
\label{fplotb}
\end{figure}
\end{center}

\begin{center}
\begin{figure}[htb]
\scalebox{0.47}{\includegraphics{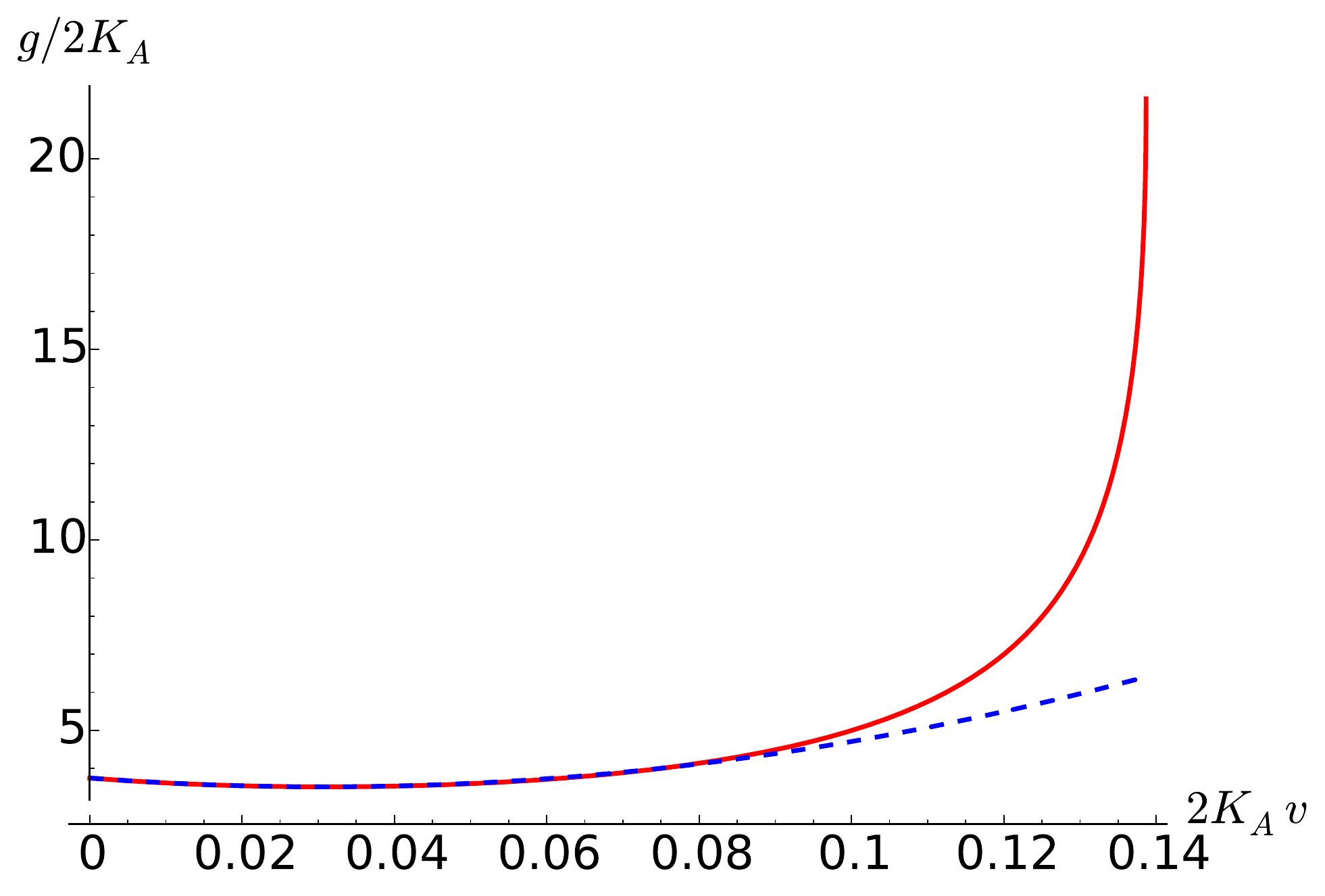}}
\caption{(Color online) Full line: the exact function $g=g\left( v\right) $; dashes: the approximation to $g$ from eq. (\ref{aproxfg}) both as a function of the invariant $2K_Av$ and scaled by the factor $2K_A$ }
\label{gplotb}
\end{figure}
\end{center}

As long as $v$ is increasing, the relationship of $v$ to $\tau$ may be inverted, and so $f$ and $g$ may be written as functions of $v$.  $f$ always starts at $f=1$ for $v=0$, while $g$ starts at a value

\be 
g_0=g\left( v=0\right) =\frac{15}{4}2K_A
\label{gnot}
\te
We may find the functions $f$ and $g$ by numerical means, the result is shown in figs. (\ref{fplotb}) and (\ref{gplotb}). If we use these functions to build the Massieu function, we obtain a theory that reproduces the free streaming solution  up to $2K_Av\approx 0.14$

If we only require agreement for early times, when the assumption of free streaming is physically plausible, then it is possible to replace the numerically found functions $f$ and $g$ by simple analytical expressions, for example

\bea
g&=&2K_A\left[3.75-15.16\left(2K_Av\right) +250\left( 2K_Av\right)^2\right]\nn
f&=&1+65\left(2K_Av\right)^2
\label{aproxfg}
\tea
Except the independent terms, the coefficients were found by trial and error. In figs. (\ref{fplotb}) and (\ref{gplotb}) we superimpose these approximations to the exact solutions from last Section. 

\section{Comparison to second order hydrodynamics}

The results from last Section suggest that it is enough to explore DTTs where the generating function $\varphi$ has no explicit dependence with respect to $ X^ 3$, $X^ 4$, $Y^ 3$ and $Y^ 4$; this is consistent with the constraints (\ref{const4}) and (\ref{const5}). More concretely, we assume

\be 
\varphi =\frac16\sigma +\sigma K_A\left\lbrace h_1\left( X^2\right) X^1+f\left( X^2\right) Y^1+h_2\left( X^2\right) +g\left( X^2\right) Y^2\right\rbrace 
\te
Then, if
\bea
g&=&\sum g_n\left(X^2\right)^{n/2}\nn
f&=&\sum f_n\left(X^2\right)^{n/2}
\tea
we find, from eqs. (\ref{const3}) and (\ref{const5})

\bea
h_2&=&\sum \frac{g_n}{2\left(n+2\right)\left(n+3\right)}\left(X^2\right)^{1+n/2}\nn
h_1&=&\sum \frac{f_n}{2\left(n+2\right)}\left(X^2\right)^{n/2}
\tea
We now consider viscosity. For a conformal theory, not all the equations (\ref{dtteq}) may be chosen independently, since they are constrained by the conditions that $v^{\mu\nu}$ be traceless and transverse. A convenient set of independent equations for the NEC is given by

\be
\left[\Delta^{\rho}_{\lambda}\Delta^{\sigma}_{\tau}-\frac13\Delta^{\rho\sigma}\Delta_{\lambda\tau}\right] A^{\mu\lambda\tau}_{;\mu}=\bar{I}^{\rho\sigma}
\te
where $\bar{I}^{\rho\sigma}$ is symmetric, traceless and transverse. Moreover the entropy production is $S^{\mu}_{;\mu}=-\zeta_{\rho\sigma}\bar{I}^{\rho\sigma}$. From the symmetries of the problem and on dimensional grounds we write

\be
\bar{I}^{\rho\sigma}=-T^6\left[\gamma_1^{-1}v^{\rho\sigma}+\gamma_3^{-1}\left(\tilde{v^3}\right)^{\rho\sigma}\right]
\te
It is interesting to compare this theory with the usual second order hydrodynamics. To this effect we only need to develop $T^{\mu\nu}$ to second order in $v$, and $A^{\mu\lambda\tau}$ to first order, namely 

\bea
T^{\mu\nu}&=&T^4t^{\mu\nu}\nn
A^{\mu\nu\rho}&=&T^5a^{\mu\nu\rho}
\tea
where

\bea
\varphi &=&\frac16\sigma +\sigma K_A\left\lbrace \left[\frac{f_0}{4}+\frac{f_2X^2}{6}\right] X^1+\left[f_0+f_2X^2\right] Y^1+\frac{g_0}{12}X^2+g_0 Y^2\right\rbrace \nn
\rho&=&\sigma+\frac 72\sigma K_Ag_0X^2\nn
t^{\mu\nu}&=&\left[ u^{\mu}u^{\nu}+\frac13\Delta^{\mu\nu}\right] \rho +2\sigma K_Af_0{v}^{\mu\nu}+2\sigma K_Ag_0\left(\tilde{v}^2\right)^{\mu\nu}\nn
a^{\mu\nu\rho}&=&\sigma K_Af_0\left[ 3u^{\mu}u^{\nu}u^{\rho}+u^{\mu}\Delta^{\nu\rho}+\Delta^{\mu\nu}u^{\rho}+\Delta^{\mu\rho}u^{\nu}\right] \nn
&+&\sigma K_A g_0 \left(v^{\mu\nu}u^{\rho}+v^{\mu\rho}u^{\nu}+v^{\nu\rho}u^{\mu}\right)
\tea
Although $T$ is not the physical temperature, the difference is not significative at this accuracy. We then find the viscous energy-momentum tensor

\be
\Pi^{\mu\nu}=2\sigma K_AT^4\left[f_0{v}^{\mu\nu}+g_0\left(\tilde{v}^2\right)^{\mu\nu}\right]
\te
or else

\be
{v}^{\mu\nu}=\frac{\Pi^{\mu\nu}}{2\sigma K_AT^4f_0}-\frac{g_0}{f_0}\frac{\left(\tilde{\Pi}^2\right)^{\mu\nu}}{\left(2\sigma K_AT^4f_0\right)^2}
\te
where

\be
\left(\tilde{\Pi}^2\right)^{\mu\nu}=\left({\Pi}^2\right)^{\mu\nu}-\frac13\Delta^{\mu\nu}\left({\Pi}^2\right)^{\rho}_{\rho}
\te
The conservation law for the nonequilibrium tensor becomes

\be
\Pi^{\tau\sigma}=-\frac{2\sigma K_Af_0\gamma_1}{T^2}\left\{\left[\Delta^{\tau}_{\mu}\Delta^{\sigma}_{\nu}-\frac13\Delta^{\tau\sigma}\Delta_{\mu\nu}\right]\left[T^5a^{\mu\nu\rho}\right]_{;\rho}\right\}+\frac{g_0}{2\sigma K_Af_0^2T^4}\left(\tilde{\Pi}^2\right)^{\tau\sigma}
\te
Expanding the derivatives we get

\bea 
\Pi^{\tau\sigma}&=&-2\left(\sigma K_Af_0\right)^2\gamma_1T^3\sigma^{\tau\sigma}\nn
&-&2\left(\sigma K_A\right)^2f_0g_0\gamma_1T^{-1}\left[\Delta^{\tau}_{\mu}\Delta^{\sigma}_{\nu}-\frac13\Delta^{\tau\sigma}\Delta_{\mu\nu}\right]\left(T^4v^{\mu\nu}\right)_{;\rho}u^{\rho}\nn
&-&2\left(\sigma K_A\right)^2f_0g_0\gamma_1T^3\left[\Delta^{\tau}_{\mu}\Delta^{\sigma}_{\nu}-\frac13\Delta^{\tau\sigma}\Delta_{\mu\nu}\right]\left[v^{\mu\nu}u^{\rho}_{;\rho}+v^{\mu\nu}u^{\rho}\left(\ln T\right)_{;\rho}+v^{\mu\rho}u^{\nu}_{;\rho}+v^{\nu\rho}u^{\mu}_{;\rho}\right]\nn
&+&\frac{g_0}{2\sigma K_Af_0^2T^4}\left(\tilde{\Pi}^2\right)^{\tau\sigma}
\tea
where

\be 
\sigma^{\tau\sigma}=\left[\Delta^{\tau\mu}\Delta^{\sigma\nu}+\Delta^{\tau\nu}\Delta^{\sigma\mu}-\frac 23\Delta^{\tau\sigma}\Delta^{\mu\nu}\right]u_{\mu;\nu}
\te
Introducing the decomposition

\be
\Delta^{\nu\rho}u^{\mu}_{;\rho}=\frac12\sigma^{\mu\nu}+\frac12\omega^{\mu\nu}+\frac13\theta\Delta^{\mu\nu}
\te
$\omega^{\mu\nu}=-\omega^{\nu\mu}$ and $\theta =\Delta^{\rho}_{\mu}u^{\mu}_{;\rho}$ and defining

\be
\dot{\Pi}^{\tau\sigma}=\left[\Delta^{\tau}_{\mu}\Delta^{\sigma}_{\nu}-\frac13\Delta^{\tau\sigma}\Delta_{\mu\nu}\right]\Pi^{\mu\nu}_{;\rho}u^{\rho}
\te
we recognize that $\eta=2\left(\sigma K_Af_0\right)^2\gamma_1T^3$ and $\tau_{\pi}=\sigma K_Ag_0\gamma_1T^{-1}$ are the usual viscosity and relaxation time, and we write

\bea 
\tau_{\pi}\dot{\Pi}^{\tau\sigma}+\Pi^{\tau\sigma}&=&-\eta\sigma^{\tau\sigma}\nn
&-&\tau_{\pi}\left[\frac 53 \Pi^{\tau\sigma} \theta + \frac12 \left(\Pi^{\tau}_{\rho} \sigma^{\sigma\rho} +\Pi^{\sigma}_{\rho}\sigma^{\tau\rho}\right)+\frac12\left(\Pi^{\tau}_{\rho}\omega^{\sigma\rho}+\Pi^{\sigma}_{\rho}\omega^{\tau\rho}\right)\right]\nn
&+&\frac{\tau_{\pi}}{\eta}\left(\tilde{\Pi}^2\right)^{\tau\sigma}-\tau_{\pi}\Pi^{\tau\sigma}u^{\rho}\left(\ln T\right)_{;\rho}
\label{finaleq}
\tea
The first two lines in eq. (\ref{finaleq}) actually reproduce eq. (24) from \cite{FJMRS15}, after allowing for the different conventions and identifying the several transport parameters, which are those from eqs (34) and (35) in that reference provided we set $J_{63}^{\left( 3\right) }=0$. The remaining terms in the third line, after noting that from the lowest order energy-momentum conservation equation we have $u^{\rho}\left(\ln T\right)_{;\rho}=\theta/3$, are seen to correspond to terms of second order in inverse Reynolds number as given in ref. \cite{DNMR12}.

\section{Final Remarks}

The main contribution of this paper is that we display the most general nonlinear hydrodynamic theory based on two conserved currents, the EMT and the NEC, having the symmetries of the second and third moments of the 1PDF, and where these currents may be derived from a vector potential.

Formally, the existence of a potential is essential to give the approach some predictive power; without it, there are just too many possibilities. The derivability from a potential ensures that there must be relations between the transport functions that appear in the conservation equation for the EMT and in the NEC. Those relations are the real predictions of the theory.

The existence of a potential, on the other hand, follows naturally if there is a non equilibrium entropy current $S^{\mu}$ and  the Second law (positivity of the entropy production $S^{\mu}_{;\mu}$) has to follow from the conservation laws at every single event. This means that we should be able to write a local relation

\be 
S^{\mu}_{;\mu}=-\beta_{\nu}T^{\mu\nu}_{;\mu}-\zeta_{\rho\sigma}A^{\mu\rho\sigma}_{;\mu}
\te
and this in turn suggests 

\be
dS^{\mu}=-\beta_{\nu}dT^{\mu\nu}-\zeta_{\rho\sigma}dA^{\mu\rho\sigma}
\te
whereby one would find the potential as a local Legendre transform of the entropy current. Both ideal hydrodynamics and first order theories do admit a potential \cite{HisLin83,HisLin85,HisLin88,Ols90}, and the equations we have found reduce to those of second order hydrodynamics near thermal solutions, albeit  for specific choices of the transport coefficients. 

A simple, sufficient condition for hydrodynamics, as derived from kinetic theory, to admit a potential, has remained elusive \cite{NaRe95}.  In general, if we assume as 1PDF a deformed Bose-Einstein distribution

\be 
f=\frac{1}{e^{-\beta_{\mu}p^{\mu}-\zeta_{\mu\nu}p^{\mu}p^{\nu}}-1}
\label{formal1}
\te 
then the second and third momenta $T^{\mu\nu}$ and $A^{\mu\nu\rho}$ may be found as in equation (\ref{potential}), where 

\be 
\Phi^{\mu}=\tau\int\frac{dp^{\chi}d^2\mathbf{p_{\perp}}}{\left(2\pi\right)^3p^0}p^{\mu}\;\ln\left[ 1+f\right] 
\label{formal2}
\te
The entropy flux eq. (\ref{entropy}) has the right form

\be 
S^{\mu}=\tau\int\frac{dp^{\chi}d^2\mathbf{p_{\perp}}}{\left(2\pi\right)^3p^0}p^{\mu}\;\left[ \left( 1+f\right) \ln\left( 1+f\right) -f\ln f\right] 
\te 
and $f$ maximizes the entropy flux for the given $T^{\mu\nu}$ and $A^{\mu\nu\rho}$ \cite{Anile}. From this point of view, it would seem that a DTT framework follows naturally if the dynamics is restricted to the manifold eq. (\ref{formal1}) in the space of distribution functions \cite{GorKar05}. However, it must be pointed out that we wish to consider tensors $\zeta_{\mu\nu}$ which are not non positive, and so for which eq. (\ref{formal1}) does not really define a 1PDF; therefore, eq. (\ref{formal2}) is at best a formal expression. For further discussion we refer the reader to refs. \cite{NaRe95} and \cite{MulRug93}

The proof that given an exact boost invariant, axisymmetric, free streaming solution there is some potential leading to currents matching the exact ones is just a tour de force to demonstrate the flexibility afforded by these theories. More interesting is the observation that theories with simple potentials already capture the early time dynamics. The point is that these theories are fully nonlinear, and so provide a natural framework to further analyze strong phenomena such as turbulence and instabilities. For these purposes, of course, it will  be necessary to expand the present framework to account for a finite relaxation time and interaction with color fields \cite{PRC12,Cal13b}. Another pending issue is to consider flows with transverse expansion \cite{DHMNS14a, DHMNS14b} and/or on curved manifolds \cite{NorDen15}

The three fields where to expect these simplified theories will prove useful are the nonlinear unfolding of plasma instabilities \cite{Mrow94,Mrow07,MaMr06,SSGT06,MrTh07,MaMa07,ARS13,AKLN14,RSA08,RS10,IRS11,ARS13b}, strong shocks \cite{OlsHis90,JoPa91,DKKM08,Bou10a,Bou10b,KKM10,KKM11,Bou14} and turbulence \cite{FlWi11,Fuk13,Khac08,CaRe10}. We expect to report soon on progress in these directions.

\begin{acknowledgments}
It is a pleasure to acknowledge exchanges with A. Kandus, J. Peralta-Ramos, J-Y Ollitrault, P. Romatschke, L. Lindblom and M. Strickland.

This work has been supported in part by ANPCyT, CONICET and University of Buenos Aires (Argentina).
\end{acknowledgments}

\section*{Appendix: Derivation of eqs. (\ref{gen1})}
We start by taking variations of the potential to get

\bea
t^{\mu\nu}&=&u^{\mu}u^{\nu}\left[6\varphi+6\sum_{a=2}^4 aX^a\frac{\partial\varphi}{\partial X^a} +\sum_{a=2}^4 4a\left(1+a\right)X^a\frac{\partial\varphi}{\partial X^a}+\sum_{a,b=2}^4 4abX^aX^b\frac{\partial^2\varphi}{\partial X^a\partial X^b}\right]\nn
&+&2\Delta^{\mu\nu}\left[\varphi+\sum_{a=2}^4 aX^a\frac{\partial\varphi}{\partial X^a}\right]+2\sum_{a=1}^4\frac{\partial\varphi}{\partial Y^a}\left( v^a\right) ^{\mu\nu}\nn
a^{\mu\nu\rho}&=&2u^{\mu}\sum_{a=1}^4\left( v^{a-1}\right) ^{\nu\rho}\left[a\left(a+1\right)\frac{\partial\varphi}{\partial X^a}+\sum_{b=2}^4abX^b\frac{\partial^2\varphi}{\partial X^a\partial X^b}\right]\nn
&+&2u^{\mu}u^{\nu}u^{\rho}\left[3\frac{\partial\varphi}{\partial Y^1}+\sum_{a=2}^4 aX^a\frac{\partial^2\varphi}{\partial X^a\partial Y^1}\right]+\sum_{a=1}^4\frac{\partial\varphi}{\partial Y^a}\left(\left( v^{a-1}\right) ^{\mu\nu}u^{\rho}+\left( v^{a-1}\right) ^{\mu\rho}u^{\nu}\right)
\tea
For $t^{\mu\nu}$ to be traceless we require

\be
0=\sum_{a=2}^4 4a\left(1+a\right)X^a\frac{\partial\varphi}{\partial X^a}+\sum_{a,b=2}^4 4abX^aX^b\frac{\partial^2\varphi}{\partial X^a\partial X^b}-2\sum_{a=2}^4\frac{\partial\varphi}{\partial Y^a}X^{a}
\label{const1}
\te
So we can write 

\be
t^{\mu\nu}=\left[ u^{\mu}u^{\nu}+\frac13\Delta^{\mu\nu}\right] \rho +2\sum_{a=1}^4\frac{\partial\varphi}{\partial Y^a}\left(\tilde{v}^a\right)^{\mu\nu}
\te
and 

\bea
a^{\mu\nu\rho}&=&2u^{\mu}g^{\nu\rho}\left[2\frac{\partial\varphi}{\partial X^1}+\sum_{b=2}^4bX^b\frac{\partial^2\varphi}{\partial X^1\partial X^b}\right]\nn
&+&2u^{\mu}\sum_{a=2}^4\left( v^{a-1}\right) ^{\nu\rho}\left[a\left(a+1\right)\frac{\partial\varphi}{\partial X^a}+\sum_{b=2}^4abX^b\frac{\partial^2\varphi}{\partial X^a\partial X^b}\right]\nn
&+&2u^{\mu}u^{\nu}u^{\rho}\left[3\frac{\partial\varphi}{\partial Y^1}+\sum_{a=2}^4 aX^a\frac{\partial^2\varphi}{\partial X^a\partial Y^1}\right]\nn
&+&\frac{\partial\varphi}{\partial Y^1}\left(g^{\mu\nu}u^{\rho}+g^{\mu\rho}u^{\nu}\right)\nn
&+&\sum_{a=2}^4\frac{\partial\varphi}{\partial Y^a}\left(\left( v^{a-1}\right) ^{\mu\nu}u^{\rho}+\left( v^{a-1}\right) ^{\mu\rho}u^{\nu}\right)
\tea
We reorder this as

\bea
a^{\mu\nu\rho}&=&2u^{\mu}u^{\nu}u^{\rho}\left[2\frac{\partial\varphi}{\partial Y^1}+\sum_{a=2}^4 aX^a\frac{\partial^2\varphi}{\partial X^a\partial Y^1}-2\frac{\partial\varphi}{\partial X^1}-\sum_{b=2}^4bX^b\frac{\partial^2\varphi}{\partial X^1\partial X^b}\right]\nn
&+&2u^{\mu}\Delta^{\nu\rho}\left[2\frac{\partial\varphi}{\partial X^1}+\sum_{b=2}^4bX^b\frac{\partial^2\varphi}{\partial X^1\partial X^b}+\sum_{a=3}^4\frac{X^{a-1}}{3} \left[a\left(a+1\right)\frac{\partial\varphi}{\partial X^a}+\sum_{b=2}^4abX^b\frac{\partial^2\varphi}{\partial X^a\partial X^b}\right]\right]\nn
&+&\left(\Delta^{\mu\nu}u^{\rho}+\Delta^{\mu\rho}u^{\nu}\right)\left[\frac{\partial\varphi}{\partial Y^1}+ \sum_{a=3}^4\frac{\partial\varphi}{\partial Y^a}\frac{X^{a-1}}{3}\right]  \nn
&+&2u^{\mu}\sum_{a=2}^4\left( \tilde{v}^{a-1}\right) ^{\nu\rho}\left[a\left(a+1\right)\frac{\partial\varphi}{\partial X^a}+\sum_{b=2}^4abX^b\frac{\partial^2\varphi}{\partial X^a\partial X^b}\right]\nn
&+&\sum_{a=2}^4\frac{\partial\varphi}{\partial Y^a}\left(\left( \tilde{v}^{a-1}\right) ^{\mu\nu}u^{\rho}+\left( \tilde{v}^{a-1}\right) ^{\mu\rho}u^{\nu}\right)
\tea
For $a^{\mu\nu\rho}$ to be traceless on the $\mu$, $\nu$ indices we need

\bea
0&=&-2\left[2\frac{\partial\varphi}{\partial Y^1}+\sum_{a=2}^4 aX^a\frac{\partial^2\varphi}{\partial X^a\partial Y^1}-2\frac{\partial\varphi}{\partial X^1}-\sum_{b=2}^4bX^b\frac{\partial^2\varphi}{\partial X^1\partial X^b}\right]\nn
&+&3\frac{\partial\varphi}{\partial Y^1}+ \sum_{a=3}^4\frac{\partial\varphi}{\partial Y^a}{X^{a-1}}
\label{const2}
\tea
For $a^{\mu\nu\rho}$ to be traceless on the $\nu$, $\rho$ indices we need

\bea
0&=&2\left[ 2\frac{\partial\varphi}{\partial X^1}+\sum_{b=2}^4bX^b\frac{\partial^2\varphi}{\partial X^1\partial X^b}+\sum_{a=3}^4\frac{X^{a-1}}{3} \left[a\left(a+1\right)\frac{\partial\varphi}{\partial X^a}+\sum_{b=2}^4abX^b\frac{\partial^2\varphi}{\partial X^a\partial X^b}\right]\right] \nn
&-&\left[\frac{\partial\varphi}{\partial Y^1}+ \sum_{a=3}^4\frac{\partial\varphi}{\partial Y^a}\frac{X^{a-1}}{3}\right]
\label{const3}
\tea
so finally

\bea
a^{\mu\nu\rho}&=&\left[ 3u^{\mu}u^{\nu}u^{\rho}+u^{\mu}\Delta^{\nu\rho}+\Delta^{\mu\nu}u^{\rho}+\Delta^{\mu\rho}u^{\nu}\right] \left[\frac{\partial\varphi}{\partial Y^1}+ \sum_{a=3}^4\frac{\partial\varphi}{\partial Y^a}\frac{X^{a-1}}{3}\right]\nn
&+&2u^{\mu}\sum_{a=2}^4\left( \tilde{v}^{a-1}\right) ^{\nu\rho}\left[a\left(a+1\right)\frac{\partial\varphi}{\partial X^a}+\sum_{b=2}^4abX^b\frac{\partial^2\varphi}{\partial X^a\partial X^b}\right]\nn
&+&\sum_{a=2}^4\frac{\partial\varphi}{\partial Y^a}\left(\left( \tilde{v}^{a-1}\right) ^{\mu\nu}u^{\rho}+\left( \tilde{v}^{a-1}\right) ^{\mu\rho}u^{\nu}\right)
\tea
On shell, $v$ is both traceless and transverse. Therefore, it can be written in some frame as

\be 
v^{\mu}_{\nu}=\left( 
\begin{array}{cccc}
0 & 0 & 0 & 0\\
0 & v_++v_- & 0 & 0\\
0 & 0 & v_+-v_- & 0 \\
0 & 0 & 0 & -2v_+
\end{array}
\right) 
\te
This implies

\bea 
\left( v^3\right) ^{\mu}_{\nu}&=&\frac13X^3\Delta^{\mu}_{\nu}+\frac12X^2v^{\mu}_{\nu}\nn
\left( v^4\right) ^{\mu}_{\nu}&=&\frac13X^3v^{\mu}_{\nu}+\frac12X^2\left( v^2\right)^{\mu}_{\nu}
\tea 
Observe that $X^2=2\left(v_-^2+3v_+^2 \right)$, $X^3=6v_+\left( v_-^2-v_+^2\right) $ and $X^4=2\left(v_-^2+3v_+^2 \right)^2$. We therefore get

\bea 
\left( \tilde{v}^3\right) ^{\mu}_{\nu}&=&\frac12X^2v^{\mu}_{\nu}\nn
\left( \tilde{v}^4\right) ^{\mu}_{\nu}&=&\frac13X^3v^{\mu}_{\nu}+\frac12X^2\left( \tilde{v}^2\right)^{\mu}_{\nu}
\tea 
and

\bea
a^{\mu\nu\rho}&=&\left[ 3u^{\mu}u^{\nu}u^{\rho}+u^{\mu}\Delta^{\nu\rho}+\Delta^{\mu\nu}u^{\rho}+\Delta^{\mu\rho}u^{\nu}\right] \left[\frac{\partial\varphi}{\partial Y^1}+ \sum_{a=3}^4\frac{\partial\varphi}{\partial Y^a}\frac{X^{a-1}}{3}\right]\nn
&+&2u^{\mu}\left( \tilde{v}\right) ^{\nu\rho}\left[6\frac{\partial\varphi}{\partial X^2}+\sum_{b=2}^42bX^b\frac{\partial^2\varphi}{\partial X^2\partial X^b}\right]+\frac{\partial\varphi}{\partial Y^2}\left(\left( \tilde{v}\right) ^{\mu\nu}u^{\rho}+\left( \tilde{v}\right) ^{\mu\rho}u^{\nu}\right)\nn
&+&2u^{\mu}\left( \tilde{v}^{2}\right) ^{\nu\rho}\left[12\frac{\partial\varphi}{\partial X^3}+\sum_{b=2}^43bX^b\frac{\partial^2\varphi}{\partial X^3\partial X^b}\right]+\frac{\partial\varphi}{\partial Y^3}\left(\left( \tilde{v}^{2}\right) ^{\mu\nu}u^{\rho}+\left( \tilde{v}^{2}\right) ^{\mu\rho}u^{\nu}\right)\nn
&+&2u^{\mu}\left( \tilde{v}^{3}\right) ^{\nu\rho}\left[20\frac{\partial\varphi}{\partial X^4}+\sum_{b=2}^44bX^b\frac{\partial^2\varphi}{\partial X^4\partial X^b}\right]+\frac{\partial\varphi}{\partial Y^4}\left(\left( \tilde{v}^{3}\right) ^{\mu\nu}u^{\rho}+\left( \tilde{v}^{a-1}\right) ^{\mu\rho}u^{\nu}\right)\nn
\tea
Imposing the symmetry conditions

\be 
24\frac{\partial\varphi}{\partial X^3}+\sum_{b=2}^46bX^b\frac{\partial^2\varphi}{\partial X^3\partial X^b}=\frac{\partial\varphi}{\partial Y^3}
\label{const4}
\te 
and

\bea
&&\frac{\partial\varphi}{\partial Y^2}+\frac12X^2\frac{\partial\varphi}{\partial Y^4}=12\frac{\partial\varphi}{\partial X^2}+\sum_{b=2}^44bX^b\frac{\partial^2\varphi}{\partial X^2\partial X^b}\nn
&+&X^2\left[20\frac{\partial\varphi}{\partial X^4}+\sum_{b=2}^44bX^b\frac{\partial^2\varphi}{\partial X^4\partial X^b}\right]
\label{const5}
\tea
we get eq. (\ref{gen1}).


\begin{thebibliography}{99}
\bibitem{BRAHMS05} BRAHMS Collaboration, 
Nucl. Phys. A757, 1 (2005).

\bibitem{PHOBOS05} PHOBOS Collaboration, 
Nucl. Phys. A757, 28 (2005).

\bibitem{STAR05} STAR Collaboration, 
Nucl. Phys. A757, 102 (2005).

\bibitem{PHENIX05} PHENIX Collaboration, 
Nucl. Phys. A757, 184 (2005).

\bibitem{Vogt07} R. Vogt,
\textit{Ultrarelativistic heavy-ion colisions
}(Elsevier, Amsterdam, 2007)

\bibitem{SSS10} S. Sarkar, H. Satz and B. Sinha (Eds.)
\textit{The Physics of the Quark-Gluon Plasma
}(Springer-Verlag, Berlin, 2010)

\bibitem{Isr88}  W. Israel, 
in A. Anile and Y. Choquet - Bruhat (eds.), \textit{Relativistic fluid dynamics} (Springer, New York, 1988).

\bibitem{BJOR83}  J.D. Bjorken, 
Phys. Rev. D {\bf 27}, 140 (1983).

\bibitem{Ruus86} P. V. Ruuskanen, 
Acta Phys. Pol. B18, 551 (1986)

\bibitem{Risc98} D. Rischke, 
Proceedings of the 11th Chris Engelbrecht Summer School in Theoretical
Physics, Cape Town, Feb. 4 - 13, 1998

\bibitem{CalHu08} E. Calzetta and B-L Hu, 
\textit{Nonequilibrium quantum field theory}
(Cambridge University Press, Cambridge (England), 2008)

\bibitem{HKB08}T. Hirano, N. van der Kolk and A. Bilandzic,
in \textit{The Physics of the Quark-Gluon Plasma}
Lecture Notes in Physics Volume 785, 2010, pp 139-178 
(ArXiv 0808.2684).

\bibitem{Roma09} P. Romatschke, 
Int.J.Mod.Phys.E19:1-53,2010 (ArXiv 0902.3663)

\bibitem{HiNa12} T. Hirano and Y. Nara, 
 Prog. Theor. Exp. Phys. 01A203 (2012)  (arXiv:1203.4418)

\bibitem{GJS13} C. Gale, S. Jeon and B. Schenke,
Int. J. of Mod. Phys. A, Vol. 28, 1340011 (2013) (arXiv:1301.5893)


\bibitem{Huo13}P. Huovinen,
Int. J. of Mod. Phys. E22 (2013) 1330029 (arXiv:1311.1849)

\bibitem{Hir14}T. Hirano,
Proceedings contribution to the International Conference on the Initial Stages of High-Energy Nuclear Collisions (IS2013)
(arXiv:1402.0913)

\bibitem{Cal13a}E. Calzetta,
Summer School on Geometric, Algebraic and Topological Methods for Quantum Field Theory, Villa de Leyva (Colombia), July 2013
(arXiv:1310.0841) 

\bibitem{JeHe15} S. Jeon and U. Heinz,
in QGP 5, edited by Xin-Nian Wang
(ArXiv:1503.03931)

\bibitem{Str13}M. Strickland,
 Invited review for Pramana
(arXiv:1312.2285)

\bibitem{Gel13} F. Gelis,
arXiv:1312.5497

\bibitem{EpGe14} T. Epelbaum and F. Gelis,
 International Conference on the Initial Stages of High-Energy Nuclear Collisions 2013 (IS2013).
(arXiv: 1401.1666)




\bibitem{Mrow94} S. Mrowczynski,
Phys. Rev. C VOLUME 49 2191 1994

\bibitem{Mrow07} S. Mrowczynski,
Nuclear Physics A 785 (2007) 128

\bibitem{MaMr06} C. Manuel and S. Mrowczynski,
Phys. Rev. D 74, 105003 (2006)

\bibitem{SSGT06}B. Schenke, M. Strickland, C. Greiner and M. H. Thoma,
Phys. Rev. D 73, 125004 (2006)

\bibitem{MrTh07} S. Mrowczynski and M. H. Thoma,
Annu. Rev. Nucl. Part. Sci. 2007. 57:61–94

\bibitem{MaMa07} M. Mannarelli and C. Manuel,
Phys. Rev. D 76, 094007 (2007)

\bibitem{ARS13}M. Attems, A. Rebhan, and M. Strickland,
Confinement X proceedings
(ArXiv:1301.7749)

\bibitem{AKLN14} M. C. A. York, A. Kurkela, E. Lu and G. D. Moore,
Phys. Rev. D 89, 074036 (2014)

\bibitem{PRC13b}E. Calzetta and J. Peralta-Ramos,
Phys. Rev. D 88, 095010 (2013)

\bibitem{OlsHis90} T. Olson and W. Hiscock, 
Ann. Phys. 204, 331 (1990)

\bibitem{JoPa91} D. Jou  and D. Pav\'on, Diego,
Phys. Rev. A44, 6496 (1991)

\bibitem{DKKM08} G.S.Denicol, T. Kodama, T. Koide, and Ph. Mota,
Phys. Rev. C78, 034901 (2008) (ArXiv:0805.1719)

\bibitem{Bou10a} I. Bouras, E. Moln\'ar, H. Niemi, Z. Xu, A. El, O. Fochler, F. Lauciello, C. Greiner, and D.H. Rischke,
J.Phys.Conf.Ser.230, 012045 (2010) (ArXiv:1004.4615)

\bibitem{Bou10b} I. Bouras, E. Moln\'r, H. Niemi, Z. Xu, A. El, O. Fochler, C. Greiner, and D.H. Rischke,
Phys. Rev. C 82, 024910 (2010) 

\bibitem{KKM10} S. Khlebnikov, M. Kruczenski and G. Michalogiorgakis,
Phys. Rev. D82, 125003 (2010) (ArXiv: 1004.3803)

\bibitem{KKM11} S. Khlebnikov, M. Kruczenski and G. Michalogiorgakis,
Journal of High Energy Physics 07, 97 (2011)


\bibitem{Bou14} I. Bouras, B. Betz, Z. Xu, and C. Greiner,
Phys. Rev. C 90, 024904 (2014)



\bibitem{FlWi11} S. Floerchinger and U. A. Wiedemann,
JHEP 11, 100 (2011)

\bibitem{Fuk13} K. Fukushima,
Phys.Rev. C89 (2014) 024907



\bibitem{Khac08}V. Khachatryan,
Nucl. Phys. A810, 109 (2008) (ArXiv:0803.1356)

\bibitem{CaRe10} M. E. Carrington and A. Rheban,
The European Physical Journal C 71, 1787 (2011)
(ArXiv:1011.0393)



\bibitem{MaSt08}M. Martinez and M. Strickland,
Phys. Rev. C 78, 034917 (2008)

\bibitem{MaSt10}M. Martinez and M. Strickland,
Phys. Rev. C 81, 024906 (2010)



\bibitem{FRS13a}W. Florkowski, R. Ryblewski and M. Strickland,
 Phys. Rev. C 88, 024903 (2013) (ArXiv:1305.7234)


\bibitem{MaSt10b}M. Martinez and  M. Strickland,
Nuclear Physics A 848 (2010) 183–197

\bibitem{MaSt11} M. Martinez and M. Strickland,
Nuclear Physics A 856 (2011) 68–87

\bibitem{Str12}M. Strickland,
Eleventh Conference on the Intersections of Particle and Nuclear Physics (CIPANP 2012)(ArXiv:1208.2626)

\bibitem{FMRS13}W. Florkowski, M. Martinez, R. Ryblewski and M. Strickland,
Xth Quark Confinement and the Hadron Spectrum (2012) (ArXiv:1301.7539)

\bibitem{FRS13b}W. Florkowski, R. Ryblewski and M. Strickland,
Nuclear Physics A 916, 249 (2013) (ArXiv:1304.0665)

\bibitem{Str14} M. Strickland
Nucl. Phys. A 926, 92 (2014)

\bibitem{NRS14} M. Nopoush, R. Ryblewski and M. Strickland, 
Phys. Rev. C 90, 014908 (2014).

\bibitem{DFRS14} G. Denicol, W. Florkowski, R. Ryblewski and M. Strickland,
Phys. Rev. C 90, 044905 (2014).

\bibitem{Str14b} M. Strickland, 
Acta Phys. Pol. B, Vol. 45, 2355 (2014)

\bibitem{HNX14} Y. Hatta, J. Noronha, and B-W Xiao,
Phys. Rev. D 89, 051702(R) (2014)

\bibitem{HNX14b} Y. Hatta, J. Noronha and B-W Xiao,
Phys. Rev. D 89, 114011 (2014)

\bibitem{NRS15} M. Nopoush, R. Ryblewski and M. Strickland, 
Phys. Rev. D 91, 045007 (2015)

\bibitem{HMX15} Y. Hatta, M. Mart\'\i nez and B-W Xiao,
ArXiv: 1502.05894

\bibitem{RSA08} Anton Rebhan, M. Strickland and M. Attems,
Phys. Rev. D 78, 045023 (2008)

\bibitem{RS10}A. Rebhan and D. Steineder,
Phys. Rev. D 81, 085044 (2010)

\bibitem{IRS11}A. Ipp, A. Rebhan and M. Strickland,
Phys. Rev. D 84, 056003 (2011)

\bibitem{ARS13b}M. Attems, A. Rebhan and M. Strickland,
Phys. Rev. D 87, 025010 (2013)


\bibitem{MNDLJG13} H. Marrochio, J. Noronha, G. S. Denicol, M. Luzum, S. Jeon and Charles Gale,
Phys. Rev. C 91, 014903 (2015)


\bibitem{GeLi90} R. Geroch and L. Lindblom, 
Phys. Rev. D 41, 1855 (1990)

\bibitem{GeLi91}  R. Geroch and L. Lindblom, 
Ann. Phys. (NY) 207, 394 (1991)

\bibitem{Cal98} E. Calzetta, 
Class. Quant. Grav. 15, 653 (1998)

\bibitem{CaTh01}  E. Calzetta and M. Thibeault, 
Phys. Rev. D 63, 103507 (2001)

\bibitem{PRC09}J. Peralta-Ramos and E. Calzetta,
Phys. Rev. D80, 126002 (2009)

\bibitem{PRC10c}J. Peralta-Ramos and E. Calzetta,
Int. J. Mod. Phys. D19, 1721 (2010)

\bibitem{PRC10b}J. Peralta-Ramos and E. Calzetta,
Phys. Rev. C82, 054905 (2010)


\bibitem{Lax73} P. Lax, 
\textit{Hyperbolic systems of conservation laws and the mathematical theory of shock waves},
(SIAM, Philadelphia (1973))

\bibitem{NaRe95} G. B. Nagy and O. A. Reula, 
J. Phys. A 30, 1695 (1997)

\bibitem{DKR10} G. S. Denicol, T. Koide, and D. H. Rischke, 
Phys. Rev. Lett. 105, 162501 (2010)

\bibitem{DMNR12}  G. S. Denicol, E. Moln\'ar, H. Niemi and D. H. Rischke, 
Eur. Phys. J. A, 48 11 (2012) 170


\bibitem{PRC10a}E. Calzetta and J. Peralta-Ramos,
Phys.Rev.D82:106003,2010

\bibitem{PRC13a}J. Peralta-Ramos and E. Calzetta,
Phys. Rev. D 87, 034003 (2013)

\bibitem{DNMR12} G. S. Denicol, H. Niemi, E. Moln\'ar and D. H. Rischke
Phys. Rev. D 85, 114047 (2012)


\bibitem{FJMRS15} W. Florkowski, A. Jaiswal, E. Maksymiuk, R. Ryblewski and M. Strickland, 
ArXiv: 1503.03226

\bibitem{RomStr03} P. Romatschke and M. Strickland,
Phys. Rev. D68, 036004 (2003)

\bibitem{HisLin83} W. Hiscock and L. Lindblom, 
Ann. Phys. 151, 466 (1983)

\bibitem{HisLin85} W. Hiscock and L. Lindblom,  
Phys. Rev. D 31, 725 (1985)

\bibitem{HisLin88} W. Hiscock and L. Lindblom,  
Contemporary Mathematics 71, 181 (1988).

\bibitem{Ols90} T. Olson, 
Ann. Phys. 199, 18 (1990).

\bibitem{Anile} A. M. Anile and O. Muscato, Phys. Rev. B 51, 16728 (1995); A. M. Anile and M. Trovato, Phys. Lett. A 230, 387 (1997); M. Trovato and P. Falsaperla, Phys. Rev. B 57, 4456
(1998)

\bibitem{GorKar05} A. N. Gorban, I. V. Karlin, \textit{Invariant Manifolds for Physical and Chemical Kinetics},
Lect. Notes Phys. 660 (Springer, Berlin Heidelberg 2004)

\bibitem{MulRug93} I. M\"uller and T. Ruggeri, \textit{Extended Thermodynamics} (Springer - Verlag, New York, 1993)



\bibitem{PRC12}J. Peralta-Ramos and E. Calzetta,
Phys. Rev. D 86, 125024 (2012)

\bibitem{Cal13b} E. Calzetta,
AIP Conf. Proc. 1578, 74 (2014) (ArXiv:1311.1845) 

\bibitem{DHMNS14a} G. S. Denicol, U. Heinz, M. Martínez, J. Noronha and M. Strickland,
Phys. Rev. Lett 113, 202301 (2014)

\bibitem{DHMNS14b} G. S. Denicol, U. Heinz, M. Martínez, J. Noronha and M. Strickland,
Phys. Rev. D90, 125026 (2014)

\bibitem{NorDen15} J. Noronha and G. S. Denicol,
ArXiv: 1502.05892




\end{thebibliography}
\end{document}